\documentclass[pra,reprint,amsmath,amssymb,aps,superscriptaddress,longbibliography]{revtex4-2}
\usepackage{xcolor}
\usepackage{graphicx}
\usepackage{dcolumn}
\usepackage{bm}
\usepackage{hyperref}

\hypersetup{%
   colorlinks = true,
   linkcolor=blue,
   citecolor=blue,
   urlcolor=blue,
   hypertexnames=false,  
}

\usepackage{graphicx, color} 
\usepackage[normalem]{ulem}          

\usepackage[utf8]{inputenc}

\usepackage{graphicx}%
\usepackage{multirow}%
\usepackage{amsmath,amssymb,amsfonts}%
\usepackage{amsthm}%
\usepackage{mathrsfs}%
\let\mathscr\mathscr
\usepackage{xcolor}%
\usepackage{textcomp}%
\usepackage{lipsum}
\usepackage{float}     
\newcommand{\extfig}{Extended Data Fig.}

\newcommand{\ExtDataFig}     {Extended Data FIG.}

\newcommand{\de}             {\mathrm{d}}  
\newcommand{\thavg}[1]       {\langle{#1}\rangle_{_T}}
\newcommand{\NLi}            {N_{\text{\tiny Li}}}
\newcommand{\NCr}            {N_{\text{\tiny Cr}}}

\newcommand{\nCrbar}
{\bar{n}}

\newcommand{\TCr}            {T_{\text{\tiny Cr}}}
\newcommand{\muLi}           {\mu_{\text{\tiny Li}}}
\newcommand{\muCr}           {\mu_{\text{\tiny Cr}}}
\newcommand{\sigmaperpCr}    {\sigma_{\perp,\text{{\tiny Cr}}}}
\newcommand{\sigmaperpLi}    {\sigma_{\perp,\text{{\tiny Li}}}}
\newcommand{\sigmaxCr}       {\sigma_{x,\text{{\tiny Cr}}}}
\newcommand{\sigmaxLi}       {\sigma_{x,\text{{\tiny Li}}}}
\newcommand{\kBoltz}         {k_{_{\text{B}}}}
\newcommand{\kB}             {\kBoltz}
\newcommand{\muB}            {\mu_{_{\text{B}}}}

\newcommand{\microsec}       {\text{µs}}
\newcommand{\micron}         {\text{µm}}

\newcommand{\mG}             {\text{mG}}
\newcommand{\XXX}            {\langle \kth / \bar{n}^{\text{\tiny 1/3}} \rangle}  
\newcommand{\Floc}           {F_{\text{loc}}}
\newcommand{\Flocone}        {F_{\text{loc,1}}}
\newcommand{\Floctwo}        {F_{\text{loc,2}}}
\newcommand{\Floceff}        {F_{\text{loc}}^*}

\newcommand{\tmax}           {t_{\text{max}}}
\newcommand{\tint}           {t_{\text{int}}}
\newcommand{\gammaTh}        {\gamma_{_T}}
\newcommand{\gammamax}       {\gamma_{_M}}
\newcommand{\kth}            {k_{_T}}

\newcommand{\Dthres}         {D_{\rm thres}}
\newcommand{\Lcoh}           {L_{\text{coh}}}

\newcommand{\mred}           {m_{\text{red}}}
\newcommand{\dtstep}         {dt_{\text{step}}}

\newcommand{\Pcoll}          {P_{\text{coll}}}

\newcommand{\Dtret}          {\Delta t_{\text{del}}}

\newcommand{\Emf}            {E_{\text{mf}}}

\begin{document}

\title{Anomalous diffusion and  localization in a disorder-free atomic mixture}


\author{S. Finelli} 
\email{stefano.finelli@unifi.it}
\affiliation{Istituto Nazionale di Ottica del Consiglio Nazionale delle Ricerche (INO-CNR), 50019 Sesto Fiorentino, Italy}
\affiliation{\mbox{Dipartimento di Fisica e Astronomia, Universit\`{a} di Firenze, 50019 Sesto Fiorentino, Italy}}

\author{B. Restivo}
\affiliation{\mbox{Dipartimento di Fisica e Astronomia, Universit\`{a} di Firenze, 50019 Sesto Fiorentino, Italy}}
\affiliation{\mbox{European laboratory for Non-Linear Spectroscopy (LENS), 50019 Sesto Fiorentino, Italy}}

\author{A. Ciamei} 
\affiliation{Istituto Nazionale di Ottica del Consiglio Nazionale delle Ricerche (INO-CNR), 50019 Sesto Fiorentino, Italy}
\affiliation{\mbox{European laboratory for Non-Linear Spectroscopy (LENS), 50019 Sesto Fiorentino, Italy}}

\author{A. Trenkwalder} 
\affiliation{Istituto Nazionale di Ottica del Consiglio Nazionale delle Ricerche (INO-CNR), 50019 Sesto Fiorentino, Italy}
\affiliation{\mbox{European laboratory for Non-Linear Spectroscopy (LENS), 50019 Sesto Fiorentino, Italy}}

\author{M. Inguscio} 
\affiliation{\mbox{European laboratory for Non-Linear Spectroscopy (LENS), 50019 Sesto Fiorentino, Italy}}

\author{D. S. Petrov}
\email{dmitry.petrov@universite-paris-saclay.fr}
\affiliation{Universit\'{e} Paris-Saclay, CNRS, LPTMS, Orsay, 91405, France}

\author{S. E. Skipetrov}
\email{sergey.skipetrov@lpmmc.cnrs.fr}
\affiliation{Universit\'{e} Grenoble Alpes, CNRS, LPMMC, Grenoble, 38000, France}

\author{M. Zaccanti} 
\affiliation{Istituto Nazionale di Ottica del Consiglio Nazionale delle Ricerche (INO-CNR), 50019 Sesto Fiorentino, Italy}
\affiliation{\mbox{European laboratory for Non-Linear Spectroscopy (LENS), 50019 Sesto Fiorentino, Italy}}

%



\begin{abstract}
The concept of random walk, in which particles or waves undergo multiple collisions with the microscopic constituents of a surrounding medium, is central to understanding diffusive transport across many research areas~\cite{boltzmann,einstein1905,LL_PhysicalKinetics, zwanzig2001, newman2010}. However, this paradigm may break down in complex systems, where quantum interference and memory effects render the particle propagation anomalous~\cite{Metzler2000}, often fostering localization~\cite{anderson1958, havlin1987}. Here we report on the observation of such anomalous dynamics in a minimal setting: an ultracold mass-imbalanced mixture of two fermionic gases in three dimensions. We release light impurities into a gas of heavier atoms and follow their evolution across different collisional regimes. Under strong interspecies interactions, by lowering the temperature we unveil a crossover from  normal diffusion~\cite{bruun2007, Sommer2011} to subdiffusion. Simultaneously, a localized fraction of the light gas emerges, displaying no discernible dynamics over hundreds of collisions.  
Our findings, incompatible with the conventional Fermi-liquid picture~\cite{landau1957,baym1991,bruun2007, Sommer2011,Enss2019}, are instead captured by a model of an atom propagating through a (quasi-)static disordered landscape of point-like scatterers~\cite{lorentz1905,Rusek2000,Skipetrov2018}. 
These  results highlight the key role of quantum interference in our mixture, which emerges as a versatile platform for exploring disorder-free localization phenomena~\cite{Smith2017,Grover2014}.
\end{abstract}

\maketitle

Understanding how a ``test particle'' -- 
an electron, a molecule, a wave, or a more complex entity -- propagates within a many-body environment is a fundamental problem spanning physics, chemistry, biology, and even social and economic 
sciences~\cite{LL_PhysicalKinetics, zwanzig2001, newman2010}.
A landmark framework addressing this question is Boltzmann’s kinetic theory and the associated concept of random walk~\cite{boltzmann, einstein1905}, which captures the stochastic particle motion through independent collisions with the microscopic constituents of the surrounding medium, leading to particle's diffusion. Remarkably, this picture also holds for a broad class of interacting fermionic systems~\cite{Ashcroft76,baym1991,zwerger}, through the quasiparticle concept of Landau’s theory of Fermi liquids~\cite{landau1957,baym1991}.
However, there exist regimes where the classical and quasiparticle-based paradigms break down: disordered electronic matter near the metal-insulator transition~\cite{mott1990} and anomalous electron transport in
graphene-based low-dimensional systems~\cite{dean} are prominent examples where
electron motion deviates from the expected Brownian behavior. In such cases, memory and quantum interference effects can lead to anomalous subdiffusive transport
and Anderson localization~\cite{anderson1958}.

Here, we report on the observation of anomalous diffusion in an ultracold, three-dimensional mixture of fermionic lithium ($^6$Li) and chromium ($^{53}$Cr) atoms~\cite{Ciamei2022}. Specifically, we investigate the axial expansion dynamics of a small sample of Li atoms, acting as light impurity particles, released into an ideal Boltzmann gas of chromium, that plays the role of a bath of heavy point-like scatterers (mass ratio $M/m=8.8$). 
By tuning Li-Cr short-range interactions through a  Feshbach resonance~\cite{Ciamei2022A}, we explore different transport regimes under various chromium temperature and density conditions.
At high temperatures, we observe a crossover from ballistic to diffusive transport upon increasing the interaction. The crossover is quantitatively reproduced by semi-classical simulations of multiple, independent Li-Cr scattering events. Under resonant conditions, we measure small diffusion coefficients of a few diffusion quanta $\hbar/m$, consistently with previous observations in  homonuclear mixtures~\cite{Sommer2011}.
In contrast to the latter, however, we unveil subdiffusive dynamics of Li atoms when the temperature is lowered and interactions are tuned near resonance.
In this anomalous regime, not reproduced by our semi-classical simulations, the mean-square size of the cloud of Li atoms grows sub-linearly in time: $\langle x^2(t) \rangle\! \propto \!t^{\alpha}$ with $\alpha\!<\! 1$.
Moreover, the lithium density distribution features a spatially localized part that exhibits no dynamics, and whose weight monotonically increases with decreasing temperature.
These observations contrast the predictions for the spin diffusion 
in homonuclear systems~\cite{Bruun2011,Goulko2013}, but are quantitatively reproduced by a model of wave propagation 
in a random Lorentz gas of point-like scatterers~\cite{Rusek2000, Skipetrov2018}.
Our study unravels a key, enhanced role of quantum interference effects in resonant heavy-light Fermi mixtures, with the heavy component acting as a quasi-static random medium for the light one, and points toward rich transport and disorder-free localization phenomena in translationally invariant systems~\cite{Smith2017,Grover2014}, solely driven by large mass imbalance.


\begin{figure*}[t!]\centering
    \includegraphics[width=2\columnwidth]{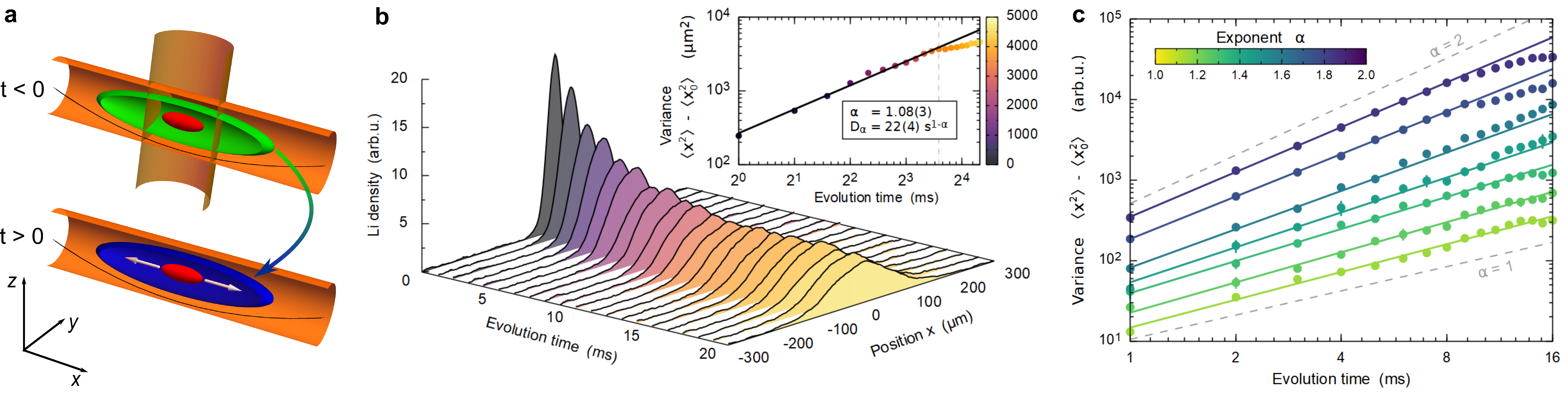}
    \vskip -5 pt
\caption{\textbf{Experimental protocol and data analysis.}
    \textbf{a}, Experimental sequence: Li$|1\rangle$ atoms (red) are initially confined at the center of a Cr$|2\rangle$ reservoir (green). A $0.9$~ms-long radio-frequency pulse maximizes the chromium transfer into the Cr$|1\rangle$ state (blue) near a Li$|1\rangle$-Cr$|1\rangle$ Feshbach resonance, and immediately after the vertical beam is switched off, allowing the Li cloud to axially expand in the presence of a weak harmonic confinement, characterized by a frequency of 17 Hz (13 Hz) for the Li (Cr) component.
    \textbf{b}, Examples of radially-integrated profiles of the expanding Li cloud. Inset: Time evolution of the corresponding mean-square displacement, which sets the color scale in the main panel.
    \textbf{c}, Examples of $\langle x^2(t) \rangle$ dynamics recorded for different $R^*/a$ values (dots), together with best fits (solid lines) to Eq.~(\ref{Eq1}), limited to $t<\tmax$ defined in the text, yielding different $\alpha$ values, see color legend. Error bars are the s.e.m. of 4–6 measurements. Data are arbitrarily shifted along the vertical direction for display purposes.}
\label{fig:fig1}
\end{figure*}

By following procedures described elsewhere~\cite{Ciamei2022,Finelli2024}, we produce weakly interacting, non-degenerate $^6$Li-$^{53}$Cr Fermi mixtures in an  optical dipole trap (ODT) oriented along the horizontal $x$-axis, see Fig.~\ref{fig:fig1}a. A magnetic field curvature provides a weak additional axial confinement. Lithium atoms are prepared in their lowest Zeeman state (denoted Li$|1\rangle$, sketched in red in Fig.~\ref{fig:fig1}a) and chromium in the second-lowest one  (Cr$|2\rangle$, sketched in green), with a typical Cr-to-Li number ratio of 10:1 and $\NCr \approx 1.2 \times 10^5$.
Application of a second, vertically-oriented ODT controllably compresses the Li axial size while  negligibly affecting the Cr cloud, of  $1/\sqrt{\text{e}}$ Gaussian axial size $\sigmaxCr$ and a typical axial-to-radial aspect ratio around 10:1. The second ODT only weakly contributes to the radial confinement of both species, with the Cr (Li)  gas featuring radial sizes of about $7\,\text{–}\,10~\micron$ ($5\,\text{–}\,8~\micron$) depending on the temperature regime.

The interspecies interaction is rapidly increased and the vertical ODT is switched off, allowing the Li cloud to axially expand through the Cr gas.
The interaction quench is realized by transferring chromium into the Cr$|1\rangle$ state (sketched in blue in Fig.~\ref{fig:fig1}a) near a narrow Li$|1\rangle$-Cr$|1\rangle$ Feshbach resonance centered at $B_0\!\approx\!1413.9$~G~\cite{Ciamei2022,Finelli2024}. The corresponding scattering amplitude is of the Breit-Wigner form: $f(k)=-(1/a+R^*k^2+ik)^{-1}$~\cite{Breit1936,chin2010}. Here $R^*=6017~a_0$ denotes the effective range parameter, $a_0$ is the Bohr radius, $\hbar k$ is the collision momentum, and $a$ is the scattering length, which varies dispersively with the magnetic field detuning $\delta B\!=\!B-B_0$~\cite{chin2010} such that $R^*/a\approx -0.3 \;\delta B /\text{mG}$ (see Methods). 
For any $k$, the scattering cross section $\sigma(k)\!=\!4\pi|f(k)|^2$, together with the local chromium density $n$, determines the mean free path $l(k)\!=\!1/(n \sigma(k))$ and scattering rate $\gamma(k)\!=\!n \sigma(k) v$, with $v$ denoting the relative velocity.

We characterize the dynamics by tracking the time evolution of the second moment of the Li distribution along the $x$ axis, $\langle x^2(t) \rangle$, obtained from best fits of the radially-integrated \textit{in situ} profiles to a generalized Gaussian envelope, see Fig.~\ref{fig:fig1}b and Methods. 
To minimize the effects of both the weak residual harmonic confinement~\cite{Uhlenbeck1930,Metzler2000} and density inhomogeneity, we restrict our analysis
to times below a maximum evolution time $\tmax$, defined either
by the ($R^*/a$-dependent) condition $\langle x^2(\tmax) \rangle=\sigmaxCr^2$, or by $\tmax = 30$~ms. 
This latter upper limit -- only relevant near resonance where the 
Li size does not reach the Cr one within 30~ms 
-- ensures a $\leq3$~mG peak-to-peak magnetic field stability, and relative atom losses~\cite{Finelli2024} below 10$\%$ level (see Methods).
With this constraint, all data are conveniently characterized, see Fig. \ref{fig:fig1}c, through a power-law fit of the form:
\begin{equation}
\langle x^2(t) \rangle = \langle x_0^2 \rangle + 2 D_\alpha t^\alpha,
\label{Eq1}
\end{equation}
where $D_\alpha$ represents a generalized diffusion constant and $\alpha$ is a scaling exponent. 
For $\alpha = 2$ and $1$, Eq.~(\ref{Eq1}) describes ballistic and diffusive transport, respectively. 
Intermediate values $1\leq \alpha \leq 2$ mark a transient ``superdiffusive'' behavior~\cite{havlin1987}, expected by fitting to Eq.~(\ref{Eq1}) the predictions for harmonically bound 
particles in diffusive media~\cite{Uhlenbeck1930} at early evolution times $t\leq 1/\gammaTh$, where $\gammaTh \!=\! \thavg{\gamma(k)}$ denotes the thermally- (and density-) averaged scattering rate. 
In contrast, $\alpha\leq1$ values are not reconcilable with this picture, pointing to a more complex scenario of anomalous subdiffusive transport~\cite{Metzler2000} beyond the independent scattering approximation (ISA).
Thus, $D_\alpha$ and $\alpha$ provide useful phenomenological measures of impurity transport under various interaction regimes~\cite{Barbosa2024}.

\begin{figure}[t!]\centering 
\includegraphics[width=\columnwidth]{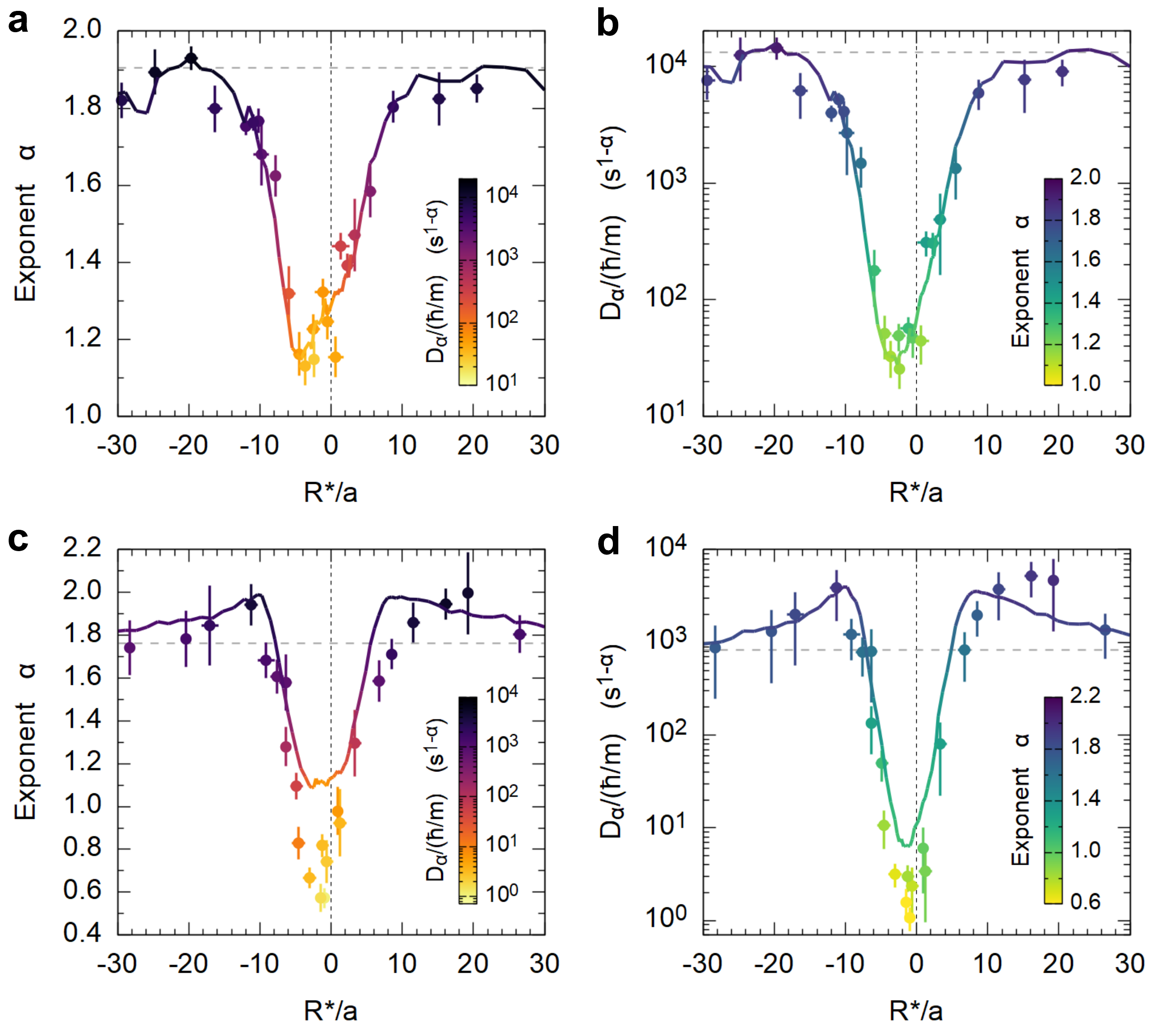}
\vskip -5 pt
\caption{\textbf{Lithium dynamics across the Li-Cr Feshbach resonance.} \textbf{a}, Scaling exponent $\alpha$ and \textbf{b}, Generalized diffusion constant $D_{\alpha}$, obtained at different $R^*/a$ values for a Cr  temperature $\TCr = 550$\,nK. 
Experimental data (dots) and semi-classical simulation predictions (solid lines) are determined from fits of $\langle x^2(t) \rangle$ to Eq.~(\ref{Eq1}), and error bars represent the fit uncertainties. 
Horizontal dashed lines mark values measured for vanishing interactions.
(\textbf{c}, \textbf{d}), Same as panels (\textbf{a}, \textbf{b}) for a lower $\TCr=350$~nK.}
\label{fig:2}
\end{figure}

Figures~\ref{fig:2}a,b (dots) respectively show the trend of the scaling exponent and generalized diffusion constant -- normalized to the quantum of diffusion $\hbar/m$ relevant for $\alpha\!=\!1$ -- as function of $R^*/a$, for a Cr gas at temperature $\TCr = 550$\,nK. As the resonance is approached and the scattering rate $\gammaTh$ increases, the dynamics crosses over from ballistic ($\alpha\!\approx\!2$) to diffusive ($\alpha\!\approx\!1$), see Fig.~\ref{fig:2}a, accompanied by a marked reduction of  $D_{\alpha}$, see Fig.~\ref{fig:2}b. 
The data are remarkably well reproduced, throughout the whole interaction crossover, by semi-classical Monte Carlo simulations based on ISA, which account for our system properties (density and momentum distributions, resonance parameters),
and which treat Li-Cr $s$-wave collisions as independent scattering events (see Methods).

The slowest dynamics is found on the negative-$a$ side of the resonance, consistent with the fact that the average scattering rate $\gammaTh$, as a function of the detuning, attains its maximum value $\gammamax$ at $R^*/a\! \approx \! -(\kth  R^*)^2 \!<$0~\cite{zaccanti2023}, where $\kth =\!\sqrt{ \kBoltz T m}/\hbar$ denotes the Li thermal wavenumber at equilibrium. For the data in Fig.~\ref{fig:2}a,b we have  $\kth \!\approx\!(7000\,a_0)^{-1}\approx 1/R^*$ explaining the shift of the minimum.
At this optimal detuning, diffusive transport is established despite the seemingly short time window explored, since $\gammamax \tmax\!\approx\!400$~\cite{Uhlenbeck1930}, see Methods. Here, we measure $D_1\!=\!15(5)~\hbar/m$, 
a value consistent with the one obtained with homonuclear $^6$Li mixtures at broad resonances under similar phase-space densities~\cite{Sommer2011,Valtolina2017}. Figures~\ref{fig:2}c,d exemplify instead the dynamics revealed at a lower Cr temperature.
 For $R^*/|a|\gg 1$
the data still match our semi-classical ISA simulations. However, near resonance our model predicts normal diffusion ($\alpha\!\approx\!1$) with reduced $D_1 < 10\,\hbar/m$, consistent with spin-transport results in the mass-balanced case~\cite{Sommer2011, Bruun2011, zaccanti2023}, 
whereas in the experiment we observe subdiffusive dynamics, with power-law exponents as low as $\alpha=0.6$ (Fig.~\ref{fig:2}c).

\begin{figure}[t]\centering 
\includegraphics[width=\columnwidth]{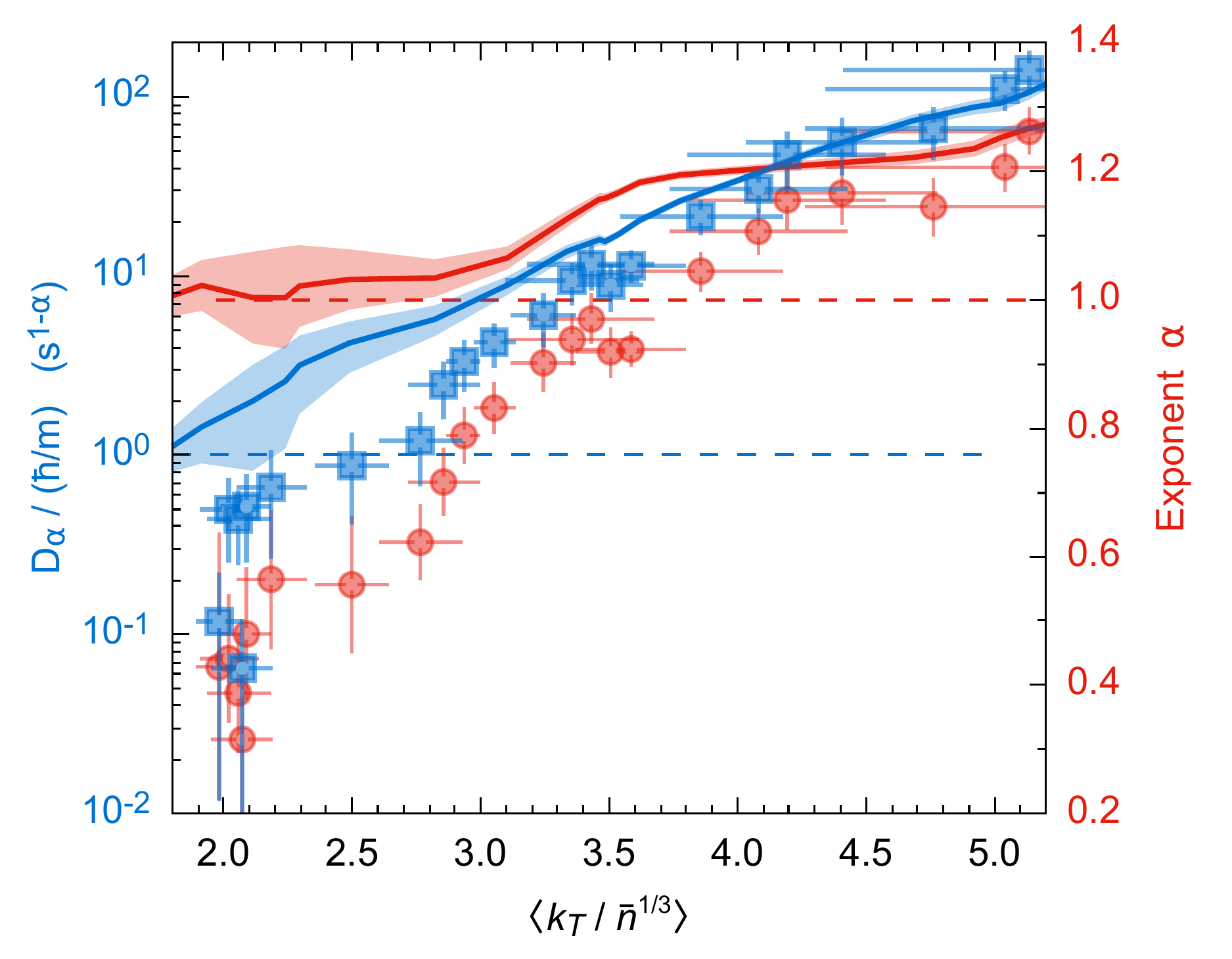}
\vskip -5 pt
\caption{\textbf{Crossover from normal to subdiffusive transport in the resonant regime.} 
Generalized diffusion coefficient $D_{\alpha}$ (blue squares, left axis) and power-law exponent $\alpha$ (red circles, right axis) as a function of the dimensionless parameter $\XXX$. 
Experimental data are compared with the results obtained from ISA simulations (solid lines with corresponding colors), based on fits to Eq.~(\ref{Eq1}). As the system transitions from normal ($\alpha \geq 1$) to subdiffusive ($\alpha < 1$) behavior, the simulation qualitatively deviates from the experimental observation. Error bars and shaded areas represent the fit uncertainties for experimental and simulated data, respectively.}
\label{fig:fig3}
\end{figure}

To gain further insight into this anomalous behavior we characterize the Li dynamics on resonance (within our experimental uncertainty we tune to $R^*/a\!=\!0$) as a function of $\XXX$, defined as the product of the Li thermal wavenumber and the Cr mean interparticle spacing, weighted on the Li density distribution and time-averaged over our observation time window, 
 see Methods. On resonance, $\XXX$ effectively quantifies the strength of interactions and the importance of multiple-scattering effects in our mixture.
The resulting $\alpha$ (red circles, right axis) and $D_{\alpha}$ (blue squares, left axis) trends are shown in Fig.~\ref{fig:fig3}. As $\XXX$ decreases, the  system crosses over from weakly superdiffusive ($\alpha > 1$), through normal diffusion ($\alpha\!\approx\! 1$), to a pronounced subdiffusive regime, where $\alpha$ values as low as 0.3 are detected. The crossover is accompanied by a strong, monotonic suppression of $D_{\alpha}$.
The weakly superdiffusive behavior at high temperatures, well captured by our ISA simulations (see solid lines in Fig.~\ref{fig:fig3}) is ascribable to residual, transient ballistic behavior ($\gammaTh \tmax\!\sim\!1$), combined with an initial temperature mismatch between Li and Cr gases. In turn, the low-temperature subdiffusion is not compatible with this picture, in light of almost instantaneous thermalization 
 observed in this regime ($\gammaTh \tmax \!\geq \!400$, see Methods), as marked by the increasing discrepancy between simulated and experimental data for $\XXX \!\lesssim\! 4$.
\begin{figure*}[t!]\centering
\includegraphics[width=1.8\columnwidth]{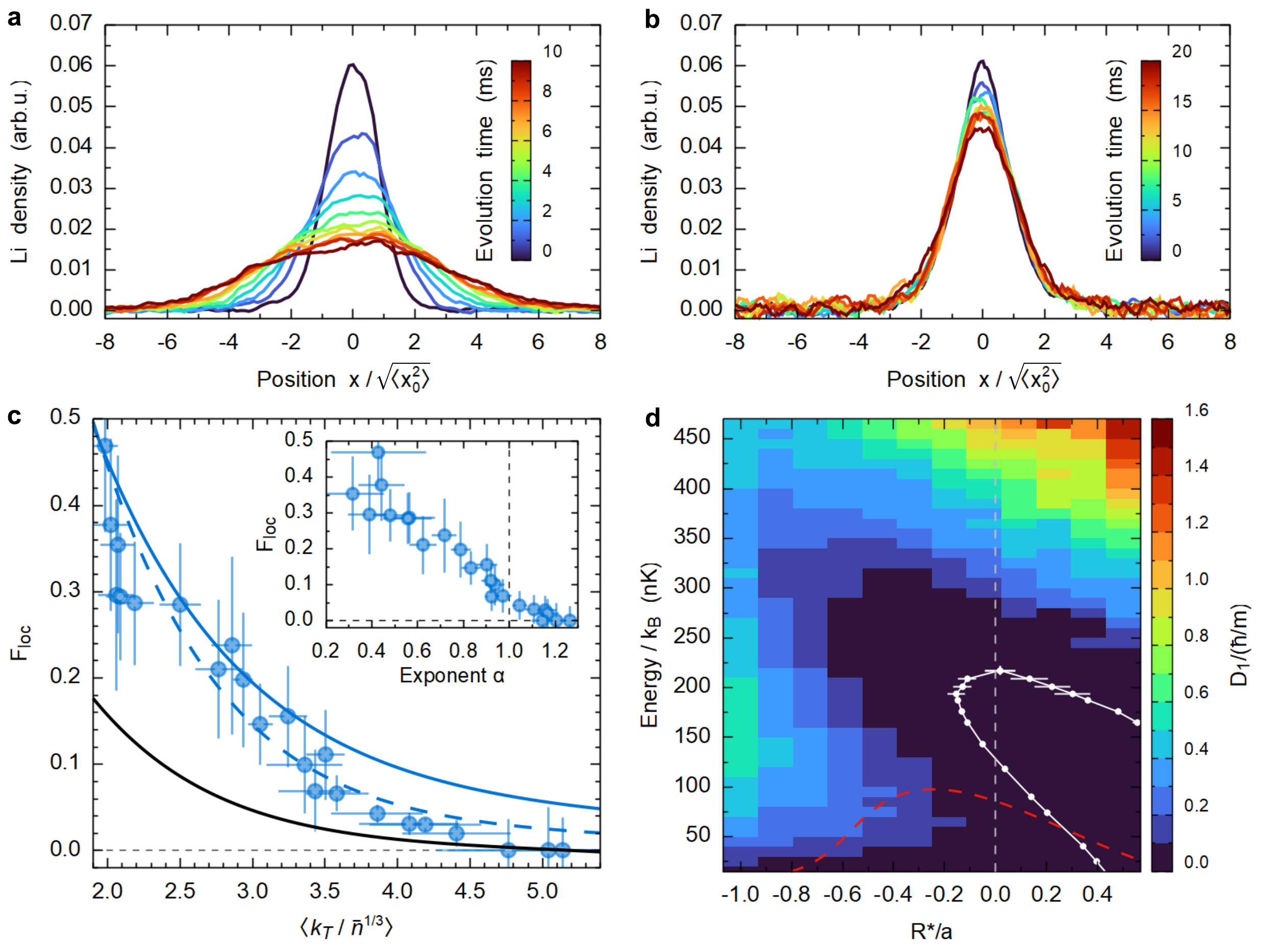}
\vskip -10pt
\caption{\textbf{Emergent Li localization in a Cr bath.} 
(\textbf{a},\textbf{b}), Normalized Li axial density profiles recorded as a function of time, see color legends, in the diffusive (\textbf{a}) and subdiffusive (\textbf{b}) regimes, respectively. In the diffusive (subdiffusive) case data are shown in 1 ms (2 ms) time steps. For each panel, the axial position is normalized to the root mean square size of the $t\!=0$ distribution. 
\textbf{c}, Experimentally-determined $\Floc$ (circles) are shown as a function of $\XXX$. Error bars combine the uncertainties of the two analyses employed to extract $\Floc$ from the density profiles (see Methods). The black line shows the predicted fraction of truly localized states of our model, 
while the blue solid line represents the total $\Floceff$ that includes also extended states with predicted $D_1\leq 0.1~\hbar/m$. 
The dashed blue line is the $\Floceff$ evaluated accounting for a finite coherence length and fluctuations of the magnetic field, see text.
\textbf{d}, Quantum diffusion constant (in $\hbar/m$ units), see color bar legend, evaluated as a function of the resonance detuning $R^*/a$ and Li energy for a random Lorentz gas with scatterer density $10^{12}$ cm$^{-3}$, is shown together with the predicted mobility edges (white dots). Red dashed line delimits the region where the independent scattering approximation predicts $D_1^{\rm ISA}<0.1~\hbar/m$.  
}
\label{fig:fig4}
\end{figure*}
We experimentally exclude that such anomalous slow-down may originate from LiCr dimer formation, and inclusion of
mean-field and retardation effects in our ISA simulations 
does not qualitatively change our simulation results (see Methods). 
We thus conclude that the subdiffusion revealed in the resonant regime is irreconcilable with a random walk picture of independent collision events. We attribute the simulation-experiment discrepancy at low $\XXX$ to enhanced quantum-interference effects in scattering of Li on multiple Cr atoms, when the lithium de Broglie wavelength becomes comparable to the mean interparticle separation of Cr atoms. It is important to note that in the mass-balanced case this condition would imply quantum degeneracy of the host component leading to the Fermi-liquid behavior and Pauli blocking of collisions. By contrast, our mixture at moderately low $\XXX$ is expected to exhibit nontrivial multiple-scattering effects without being quantum degenerate.


The crossover from normal to anomalous dynamics of $\langle x^2(t) \rangle$ 
is accompanied by a qualitative change of the evolution of the whole Li density profiles, exemplified by the raw data shown in Fig.~\ref{fig:fig4}a,b. 
In the diffusive regime, see Fig.~\ref{fig:fig4}a, the entire envelope expands remaining Gaussian at all times, as expected~\cite{Metzler2000}. 
In the anomalous region shown in Fig.~\ref{fig:fig4}b, instead, the cloud exhibits a slow (subdiffusive) drop of the central part, 
superimposed to another portion of the cloud that exhibits no detectable dynamics over the whole $(0; \tmax)$ window. This part of the Li distribution, rapidly established after the switch off of the crossed trap, is thus effectively localized, at least over our  observation timescale.
We quantify the emergence of such localized fraction $\Floc$ of the lithium distribution through two complementary analyses of the density profiles (see Methods) and the resulting trend of $\Floc$ as a function of $\XXX$ is shown in Fig.~\ref{fig:fig4}c.
Plotting $\Floc$ versus the generalized diffusion exponent $\alpha$ (Fig.~\ref{fig:fig4}c, inset) reveals a strong correlation of these quantities, showing that a non-zero localized fraction arises only for $\alpha < 1$. 
Furthermore, $\Floc$ monotonically increases with decreasing $\XXX$, reaching values as high as 40$\%$ at our lowest temperatures and highest Cr densities.

A localized component of Li gas and anomalous subdiffusive transport, persisting over hundreds of scattering times, mark a qualitative departure from both theoretical expectations and previous studies of mass-balanced Fermi mixtures~\cite{Sommer2011,Bruun2011,Valtolina2017}. By contrast, our findings bear strong similarities with the behavior of matter and sound waves in disordered potentials near localization transitions~\cite{Billy2008,Roati2008,Chabe2008,Hu2008,Semeghini2015,Cobus2018,Barbosa2024}, arising from multiple scattering and quantum interference effects. 
The mass imbalance of our  mixture leads to significantly different lithium and chromium thermal velocities, similar to unequal hopping rates in imbalanced Fermi-Hubbard models~\cite{Grover2014,Oppong2022}. 
In contrast to the mass-balanced case, a Li atom can establish coherence by scattering off and making a few return trips from one Cr scatterer to another, before thermal motion significantly alters the random configuration of chromium particles.
To leading order in $m/M$, our system can thus be mapped onto the model of a wavepacket propagating through a random Lorentz gas~\cite{lorentz1905} of resonant immobile point-like scatterers. 
This model is known to support a band of Anderson-localized states~\cite{Skipetrov2018}, separated from extended ones by a mobility edge, shown in Fig.~\ref{fig:fig4}d by white dots (see Methods).
Additionally, the color map in Fig.~\ref{fig:fig4}d shows the diffusion constant $D_1$ calculated beyond ISA (see Methods). These results demonstrate that, due to quantum interferences, $D_1$ can be strongly suppressed compared to the ISA prediction $D_1^{\rm ISA}=(\hbar/m) k\, l(k)/3$~\cite{Lee1985,Kramer1993}.
Note the difference between the region of $D_1^{\rm ISA}< 0.1\; \hbar/m$ (below the dashed red curve in Fig.~\ref{fig:fig4}d) and the region $D_1 < 0.1\;\hbar/m$ (the darkest blue area). 
The threshold  $\Dthres=0.1\; \hbar/m$ chosen for this comparison is experimentally relevant since
a diffusive propagation with $D_1 \leq \Dthres$ results in a sub-micron increase of $\sqrt{\langle x^2(t)\rangle}$ at our longest time $\tmax$, below our experimental resolution. Therefore, this slowly diffusive dynamics appears as {\it effectively} localized.

To interpret the observed $\Floc$ within the framework of the random Lorentz gas model, we assume a Li Boltzmann distribution at the equilibrium temperature $T$ and count the fraction of states with either $D_1=0$ or $D_1 < \Dthres$  (see Methods), thus obtaining theoretical predictions for the truly localized fraction $\Floc$ and the effectively localized one $\Floceff$.  These functions, obtained for $R^*/a=0$ and $\Dthres = 0.1\;\hbar/m$, are shown in Fig.~\ref{fig:fig4}c by the solid black and  blue lines, respectively. From our model, the fraction of truly localized states $\Floc$ is found to amount to about half of the observed one.
Whereas $\Floc$ 
is sensitive to variations of the detuning, we have checked that $\Floceff$ is practically insensitive to changes in $\Dthres$ within experimentally reasonable bounds: multiplying or dividing $\Dthres$ by 2 leads to a variation of $\Floceff$ well within our experimental error bars. 

From Fig.~\ref{fig:fig4}c one sees that the theoretical $\Floceff$ quantitatively matches the measured localized fraction for $\XXX \lesssim 3$, while overestimating it at higher temperatures. This discrepancy can be due to thermal decoherence effects~\cite{Lee1985}, neglected in our calculations. Interferences among different paths are indeed expected to be washed out when Cr atoms move a distance comparable to the lithium de Broglie wavelength. Multiplying the corresponding coherence time by the Li velocity, one can estimate the coherence length in our system as $\Lcoh \sim \kth^{-1}\sqrt{M/m}$. This length sets a cutoff beyond which quantum interference and localization effects are suppressed. 
 In particular, one expects $D_1$ to approach $D_1^{\rm ISA}$ once the Li mean free path $l(k)$ becomes comparable to $\Lcoh$~\cite{Kramer1993}. To account for Cr thermal motion in our $\Floceff$ calculation we thus retain only states with $D_1(k)\leq \Dthres$ and $l(k)>\delta \Lcoh$, $\delta$ being a constant of order unity, used here as a fitting parameter. 
 The resulting $\Floceff$,  shown by the dashed blue line in Fig.~\ref{fig:fig4}c for $\delta=0.6$, quantitatively reproduces the experimental trend. 
The reduction of $\Floceff$ can also  be ascribed to magnetic field fluctuations during the observation time. The averaged $\Floceff$ value, calculated for $\Lcoh=\infty$ but assuming $R^*/a$ to uniformly vary from $-0.6$ to 0.6, would lead to a result practically indistinguishable from the dashed blue curve. Our  $\leq3$~mG peak-to-peak fluctuations translate into an interval around resonance of $-0.4\leq R^*/a\leq0.4$. 
Therefore, the discrepancy between  blue solid curve and experimental data likely arises from a combination of these two effects, that are hard to disentangle through a fit, since $\delta$ and the fluctuation amplitude are strongly correlated.

The observed agreement is quite remarkable in view of the relative simplicity of our one-body theoretical framework in which Cr atoms are fixed and Li atoms are treated as a coherent matter wavepacket with a thermal energy distribution and a finite coherence length. 
We remark that our model can also explain the non-Gaussian Li envelopes in Fig.~\ref{fig:fig4}b, resulting from the evolution of a thermal ensemble of states that share the same initial spatial distribution,  but exhibit a strongly energy-dependent diffusion constant. On the other hand, the origin of subdiffusion in Fig.~\ref{fig:fig3} requires to go beyond the present framework. The full problem,  where both Li and Cr components are treated on an equal footing~\cite{Grover2014,Oppong2022}, is much more complex, and falls in the highly active research area of many-body localization~\cite{Basko2006,Abanin2019}. Understanding the full many-body dynamics of our three-dimensional, bulk system appears at present a formidable theoretical challenge  beyond the scope of this work.

In conclusion, by exploring the transport properties of a thermal lithium cloud released within an ideal gas of heavier
chromium atoms, we have unveiled, under resonantly-interacting conditions, anomalous subdiffusive dynamics, accompanied by the emergence of non-trivial localization. Our findings mark a qualitative departure from textbook expectations  based on the picture of successive but independent scattering events, at the heart of normal diffusive transport~\cite{LL_PhysicalKinetics}. The overall agreement between our experimental results and predictions by a model for multiple scattering of matter waves within a random Lorentz gas of resonant scatterers points to a key, enhanced role of quantum interference in our mixture, that suppresses diffusion and fosters lithium localization. According to our theory interpretation, this is made possible by the resonant nature of interactions in our system, which reduces both Li mean-free paths and localization lengths down to the scale of the Cr interparticle distance, so that quantum interference can persist despite the limited 
coherence length set by thermal fluctuations of the chromium gas. In particular, the large Li-Cr mass asymmetry allows to access a regime, absent for equal-mass mixtures, where both components are thermal and not affected by the Pauli suppression of collisions, but in which the de Broglie wavelength of the light atoms exceeds the mean separation between the heavy particles.
We thus foresee the relevance, and challenge, of including such effects in many-body theories aiming to describe mass- and population-imbalanced Fermi mixtures.

It will be interesting to further explore the effect of the mass imbalance, either by extending our protocols to other atomic mixtures, or by tuning the chromium (and lithium) effective mass through species-selective optical lattices~\cite{Massignan2006}. In the lattice case, by tuning the Cr filling fraction one could investigate the transition from an Anderson insulator at low fillings, to metallic or band-insulating states at unit filling, when the Cr component acts as a perfect crystal of point-like scatterers. Although in the present work the Cr bath is thermal and the resulting disorder is uncorrelated, interesting effects may arise from pair- and higher-order correlations in the heavy component becoming quantum degenerate. Such correlations in Fermi and Bose gases~\cite{Yao2025,deJongh2025,Xiang2025}, setting local constraints on the particles’ arrangement as for ``hyper-uniform'' matter~\cite{Torquato2018}, could either enhance or suppress localization of the light particles.
Another compelling possibility is enabled by the tunability of the relative Cr and Li populations, that could allow to cross over from Anderson localization~\cite{anderson1958} to Anderson orthogonality catastrophe~\cite{Anderson1967} within a single experimental setup. 
Finally,
our findings also have potential implications for strongly interacting mixtures beyond ultracold gases such as, for instance, exciton-electron matter in atomically thin semiconductors~\cite{Fey2020}.


\section*{Acknowledgements}

We gratefully acknowledge fruitful discussions with C.~Baroni, M.~Caldara, E.~Demler, T.~Enss, T.~Giamarchi, D.~Hernandez-Rajkov, M.~Kreyer, L.~Pattelli, E.~Pini, and C.~Texier.

\section*{Declarations}

\noindent\textbf{Funding}: This work was supported by the European Research Council under Grant No.~637738, by the Italian Ministry of University and Research under the PRIN2022 project No.~20227F5W4N and, co-funded by the European Union — NextGenerationEU, under the ‘Integrated infrastructure initiative in Photonic and Quantum Sciences’ I-PHOQS (CUP B53C22001750006), the PE0000023-NQSTI project, and the Young Researcher Grants MSCA\_0000042 (PoPaMol, fellowship to A.C.). 
D.S. and M.Z. acknowledge the CNR STM 2025 Program.\\
\textbf{Competing interests}: The authors declare no competing interests.\\
\textbf{Data and code availability}: All data of the figures in the manuscript and Methods, together with the most relevant codes, are available in a Zenodo repository at 10.5281/zenodo.18302590.\\
\textbf{Author contribution}:
A.C., S.F., B.R., A.T. and M.Z. performed the experiments and analyzed the data. D.P. and S.S. developed the theoretical model for wave scattering. S.F. carried out the semi-classical simulations. S.F., D.P., S.S. and M.Z. performed the experiment–theory comparison. All authors contributed to the interpretation of the results and the writing of the paper.

\noindent

\setcounter{equation}{1}  
\setcounter{figure}{0}

\section*{Methods}\label{Methods}

\subsection*{Sample preparation}\label{ExpDetail}
We produce  Li-Cr  mixtures by evaporatively cooling lithium atoms equally populating the Li$|1\rangle$ and Li$|2\rangle$  states and, simultaneously,  by sympathetically cooling  polarized Cr$|1\rangle$ atoms~\cite{Ciamei2022,Finelli2024}. 
After evaporation, the field detuning is set to $\delta B \simeq 150~\mG$ above the Li$|1\rangle$-Cr$|1\rangle$ Feshbach resonance~\cite{Ciamei2022A}, and the samples are held in the main horizontal ODT, see Fig.~\ref{fig:fig1}a, generated by a multi-mode laser beam around 1070 nm, focused down to waists $w_y=58~\micron$ and $w_z=45~\micron$. 
Gravitational sag is canceled by a magnetic gradient of about 1.5~G/cm, almost perfectly levitating chromium (magnetic moment $\muCr = 6~\muB$, with $\muB$ the Bohr magneton) while slightly over-levitating lithium ($\muLi = 1~\muB$). 
Axial confinement, provided by our magnetic-field curvature, results in a Li (Cr) harmonic frequency of about 17~Hz (13~Hz) for fields around 1400 G. Since $\muLi = \muCr/6$, thermalized mixtures feature 1/$\sqrt{\text{e}}$ Gaussian sizes in a ratio $\sigmaxLi/\sigmaxCr \!=\! \sqrt{6}$. 
To increase the Cr size, and correspondingly $\tmax$, we modify the evaporation ramps to obtain temperature-imbalanced mixtures, with the Cr effective axial temperature exceeding the Li one, in a configuration nearly stationary over a few seconds. This is made possible by the large trap aspect ratio and by the distinct scattering lengths characterizing  Li$|1\rangle$-Li$|2\rangle$ and Li-Cr thermalization rates at such fields: $a_{\mathrm{Li\!-\!Li}}\!\sim\! -2500~a_0$~\cite{Zurn2013} and $a_{\mathrm{Li\!-\!Cr}}\!\sim\!-90~a_0$~\cite{Ciamei2022A}.
A second, vertically oriented infrared laser beam at 1550~nm, focused on the magnetic curvature minimum (waists of about $100~\micron$) and only weakly contributing to the radial confinement of both species, is employed to compress the Li cloud, see Fig.~\ref{fig:fig1}a. The Cr gas is nearly unaffected by this second ODT, owing to a Li-to-Cr polarizability ratio of about  3 to 1 at such wavelength, and to a chromium temperature exceeding the Li one. 
After rising the vertical ODT, we transfer chromium in the Cr$|2\rangle$ state via a radio-frequency pulse, and subsequently remove the Li$|2\rangle$ component by an optical blast. We then lower the vertical trap power to the target value chosen for the measurement.
The resulting mixtures contain about $1.2 \!\times \!10^5$ Cr$|2\rangle$ and $10^4$ Li$|1\rangle$ atoms, and the clouds are characterized by \textit{in situ} Gaussian sizes $\sigmaxCr\sim(80; 150)~\micron$, $\sigmaperpCr\!\sim(8; 10)~\micron$, $\sigmaxLi\sim(15; 40)~\micron$, $\sigmaperpLi\!\sim(5; 7)~\micron$, 
depending on the regime experimentally explored. 
The corresponding (axial and transverse) effective temperatures are determined from time-of-flight measurements on weakly-interacting samples and their mean value, weighted with the atom numbers, defines the mixture equilibrium temperature $T$.

We turn on Li-Cr interactions at each $\delta B$ value by applying a second, 0.9 ms-long, radio-frequency pulse that maximizes the chromium transfer back into the resonant Cr$|1\rangle$ state. 
Right after, the vertical ODT is switched off and the Li gas axially expands while being negligibly excited along the transverse directions.
The system dynamics is monitored by \textit{in situ} imaging of Li$|1\rangle$ and Cr$|1\rangle$ gases, acquired on the same CCD camera.   
During each experimental run, the bias field is stabilized following  protocols described in Ref.~\cite{Finelli2024}, yielding a residual shot-to-shot noise  characterized by a standard deviation of 2.5 mG. 
Slow thermal drifts of the magnetic field are monitored by performing radio-frequency spectroscopy on  the Li$|1\rangle\leftrightarrow$Li$|2\rangle$ transition before and after each data-set acquisition: All measurements where the field drift exceeded the shot-to-shot noise level were disregarded. 
The condition $\tmax\leq30$~ms is an empirical compromise between recording the dynamics for sufficient time and with enough statistics, while keeping the probability for field drifts to occur during the measurement time around 10$\%$. Such constraint also guarantees that atom losses due to enhanced inelastic scattering for $|\delta B| \rightarrow 0$~\cite{Finelli2024} do not exceed 10$\%$.

\subsection*{Data analysis}\label{analysis}

We extract $\langle x^2(t) \rangle$ of the Li distribution by fitting radially-integrated \textit{in situ} density profiles acquired at various evolution times, see Fig.~\ref{fig:fig1}b, to a phenomenological generalized  Gaussian envelope:
\begin{equation}\label{stretchedG}
	n_{\rm Li}(x,t)=A\,\cdot\,\exp\left[\,-\frac{1}{2} \left(\frac{|x-x_0|}{\sigma_{\beta}(t)}\right)^{\beta}\,\right].  
\end{equation}
From the best-fit values of $\beta$ and $\sigma_{\beta}(t)$, $\langle x^2(t) \rangle$ is evaluated as:
\begin{equation}\label{x2_stretchedG}
\langle x^2(t) \rangle= 2^{\frac{2}{\beta}} \sigma_{\beta}^2(t) \frac{\Gamma[\frac{3}{\beta}]}{\beta  \Gamma[\frac{\beta +1}{\beta}]},	
\end{equation}
with $\Gamma[z]$ denoting the Euler Gamma function, and the corresponding error is obtained by propagating the fit uncertainties of $\sigma_{\beta}(t)$ and $\beta$. 
The results of  a few (from 4 to 6) independent measurements at fixed time and detuning are combined to obtain the mean value and s.e.m.~of $\langle x^2(t) \rangle$, based on which we characterize the Li dynamics. 
 
\begin{figure}[h!]\centering
   \includegraphics[width=\columnwidth]{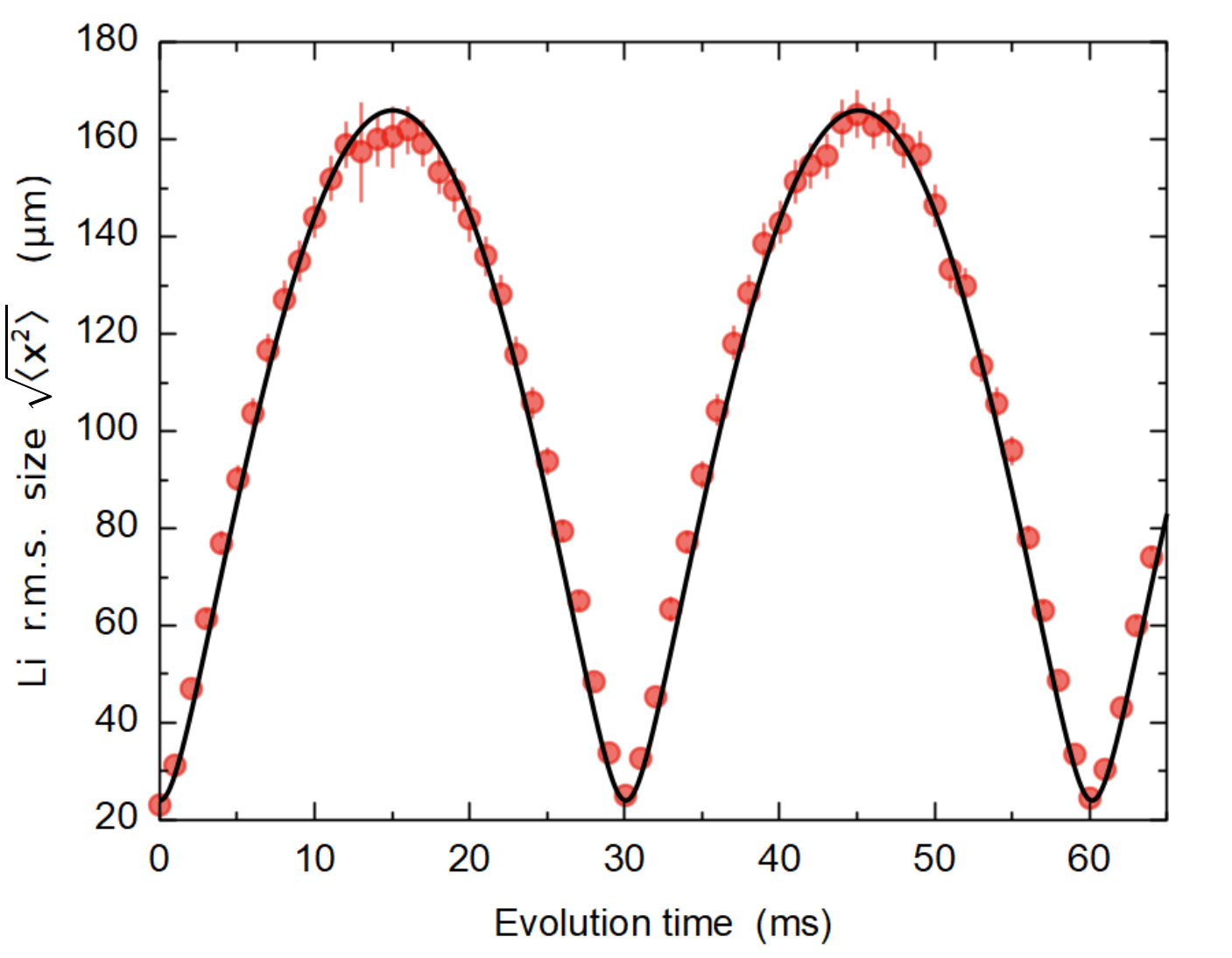}
  \caption{\textbf{Axial breathing dynamics of a non-interacting lithium cloud after release.} Dots are experimental points, error bars are the s.e.m.~of 4–6 measurements. The solid line is the best-fit based on the analytic predictions of Ref.~\cite{Uhlenbeck1930} for the non-interacting case, see also Ref.~\cite{ketterle2008}.} 
  \label{Extfig1}
\end{figure}

In the absence of  interactions ($\gamma(k)=0$) the thermal Li clouds have Gaussian shapes ($\beta$=2) of size $\sigmaxLi(t)$, and $\langle x^2(t) \rangle=\sigmaxLi^2(t)$. 
Owing to the weak, axial harmonic confinement,  $\sigmaxLi(t)$, rather than ballistically expanding, undergoes undamped breathing oscillations, see 
\extfig~\ref{Extfig1}, occurring at twice the final trap frequency $\omega_f\!= 2\pi \times 17$~Hz. 
Specifically, $\sigmaxLi(t)$ oscillates between the initial \textit{in situ} size $\sigma_{x,0}=\sqrt{\kB T_0/(m \omega_0^2)}$ and the maximum value $\sigma_{x,0}(\omega_0/\omega_f)$~\cite{ketterle2008}, in turn yielding -- up to a multiplicative factor $\hbar/(m\omega_f)$ -- the size of the initial momentum distribution, and thus $T_0$.

Analytic results are available also to describe the $\langle x^2(t) \rangle$ evolution of harmonically trapped particles within both diffusive~\cite{Uhlenbeck1930} and subdiffusive~\cite{Metzler2000} homogeneous media, throughout the crossover from the short- ($\gammaTh t\ll 1$) to long-  ($\gammaTh t\gg 1$) time limits.
Yet, these results are hard to exploit in our system, owing to the inhomogeneous Gaussian profile of the Cr bath and to the fact that, at equilibrium, $\sigmaxLi/\sigmaxCr = \sqrt{6}$. 
We thus limit our analysis to the early-time dynamics $t\leq \tmax$, such that $\langle x^2(t) \rangle$ is conveniently fitted to Eq.~(\ref{Eq1}). 
In practice, $\alpha$ and $D_{\alpha}$ are respectively obtained as best-fitted slope and intercept of a linear function to the $\mathrm{Log}[(\langle x^2(t) \rangle-\langle x_0^2 \rangle)/\micron^2]$ data, plotted versus $\mathrm{Log}[t/\text{ms}]$. The resulting intercept is then converted into the normalized diffusion coefficient $D_{\alpha}/(\hbar/m)$ (units of $s^{1-\alpha}$).

From the fits of the Cr and Li distributions at each time, we also evaluate the Cr peak density and the density-weighted Cr density $\nCrbar$. These quantities, time-averaged over the window $(0; \tmax)$, are used as inputs for our Lorentz gas model. They also define, in combination with the additional temperature measurements, the parameter $\XXX$. 

\begin{figure*}[t!]\centering
   \includegraphics[width=1.8 \columnwidth]{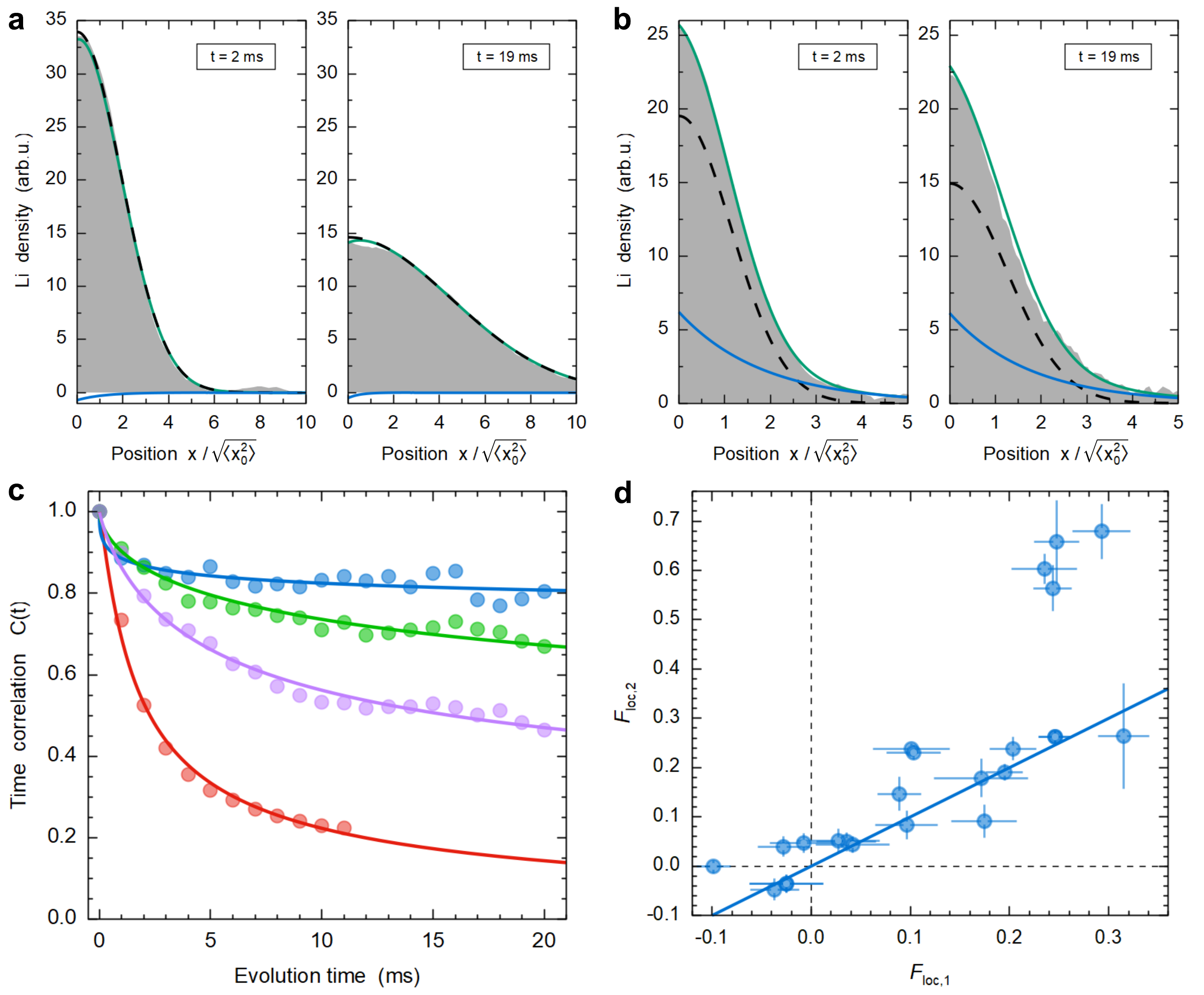}
  \caption{\textbf{Experimental methods to extract $\Floc$.}
  \textbf{a}, Examples of radially-integrated Li profiles (shaded areas) at short and long times for normally-diffusive data. The green curve is the best fit according to Eq.~(\ref{FLoc1}), with the dashed black (solid blue) line representing the Gaussian (exponential) part.
  \textbf{b}, Same as \textbf{a}, but for subdiffusive data. The exponential part shows no dynamics over the experimental timescale.
  \textbf{c}, Examples of diffusive (red dots) and subdiffusive (violet, green, blue) time-correlation data $C(t)$, together with best fits (solid lines) according to Eq.~(\ref{FLoc2}).
  \textbf{d}, Comparison between $\Flocone$ and $\Floctwo$. 
  The blue line is $y=x$.  }
  \label{Extfig_FLoc}
\end{figure*}

The determination of $\Floc$ shown in Fig.~\ref{fig:fig4}c is performed through two complementary analyses of the lithium density profiles.
The first one employs a phenomenological bimodal fitting function of the kind:
\begin{equation}\label{FLoc1}
f(x)=A \cdot \left( \frac{1-F}{\sqrt{2 \pi }\sigma_G}e^{\frac{-x^2}{2 \sigma_G^2}}+\frac{F}{4 \sigma_E}e^{\frac{-|x|}{2 \sigma_E}}\right),
\end{equation}
that, up to an overall amplitude $A$, represents a weighted sum of normalized Gaussian and exponential functions. 
We remark that both Eqs.~(\ref{stretchedG}) and~(\ref{FLoc1}) fits to the data yield consistent $\langle x^2(t)\rangle$ values, although Eq.~(\ref{stretchedG}) is preferred for that analysis given the smaller number of fitting parameters.
Normally-diffusing Li clouds are, as expected~\cite{Metzler2000}, well fitted by single Gaussian profiles both at short and long evolution times, see \extfig~\ref{Extfig_FLoc}a, correspondingly yielding, when fitted to Eq.~(\ref{FLoc1}), $F\!=\!0$ within fit uncertainty. 
This is no longer the case in the anomalous regime, see \extfig~\ref{Extfig_FLoc}b, and the Li profiles are better described by either the stretched Gaussian Eq.~(\ref{stretchedG}) with $\beta\!<\!2$, or by Eq.~(\ref{FLoc1}) with $F\!>\!0$. 
Employing Eq.~(\ref{FLoc1}) to fit the Li envelopes we find that, while the Gaussian part of the density profile evolves subdiffusively, the exponential part exhibits no detectable dynamics (neither in weight $F$ nor in size $\sigma_E$) over our observation time. We thus define the localized fraction $\Flocone$ as the time-averaged value of the fitted $F$ over the observation window $(0; \tmax)$. 

In the second method we first evaluate, from the normalized raw density profiles at each time, the time-correlation function defined as:
\begin{equation}\label{CorrFun}
C(t)=\frac{\sum_i n_{\rm Li}(i,t)n_{\rm Li}(i,0)}{\sum_i n_{\rm Li}(i,0)^2},
\end{equation}
with the sum running over all pixels of the image.
We then fit the $C(t)$ evolution to:
\begin{equation}\label{FLoc2}
g(t)=\left[\frac{2}{1+(1+2 D_{\alpha}t^{\alpha}/\langle x_0^2\rangle)^{\beta/2}}\right]^{1/\beta}(1-\Floctwo)+\Floctwo
\end{equation}
which, for $\Floctwo\!=\!0$, represents the trend of the time-correlation function of a stretched Gaussian envelope (of generalized exponent $\beta$) expanding subdiffusively with given $\alpha$ and $D_{\alpha}$ values.
As long as no part of the distribution is localized, $C(t)$ vanishes as $t\!\rightarrow\!\infty$, whereas it saturates to a non-zero offset otherwise, see examples in \extfig~\ref{Extfig_FLoc}c.
We then fit the determined $C(t)$ traces to Eq.~(\ref{FLoc2}), where $\beta$, $\alpha$ and $D_{\alpha}$ are set to the values previously determined from the $\langle x^2 \rangle$ analysis, while the offset $\Floctwo$, together with the initial second moment of the distribution $\langle x_0^2 \rangle$, are left as free fitting parameters. The resulting best-fit of $\Floctwo$ provides an alternative measure of the localized fraction.
These two rather different analyses yield consistent results for most of the regimes explored experimentally, see \extfig~\ref{Extfig_FLoc}d, the second method returning generally smaller fit uncertainties, and at our lowest $\XXX$ values, significantly higher localized fraction values. In Fig.~\ref{fig:fig4}c, we plot the mean value $(\Flocone\!+\!\Floctwo)\!/2$, and the error bar is defined as the maximum between the highest fit uncertainty and $|\Flocone\!-\!\Floctwo|/2$.

\subsection*{Semiclassical simulations}\label{Simulator}

As a first attempt to model our experiment, we developed a semi-classical Monte Carlo simulator similar to those used in Refs.~\cite{Wade2011,Sykes2015}, but tailored to our experimental conditions. 
Numerical simulations are performed under the following assumptions:
First, since in the experiment $\NLi \ll \NCr$, the lithium cloud is considered a statistical ensemble of (thermal) impurities that do not cause any effect on the larger chromium bath, treated within the local-density approximation.
Second, the external (optical and magnetic) confinement is included through the harmonic approximation: 
This implies that, in the absence of collisions, the unperturbed motion of lithium atoms within the trap has an exact analytical solution, which does not require numerical integration of the equations of motion.
Third, interactions are incorporated as (random and independent) pairwise elastic $s$-wave collisions, with detailed knowledge of the parameters characterizing the exploited Feshbach resonance~\cite{Ciamei2022A}.
Lastly, both Li and Cr atoms are always assumed to be  particles  with classical trajectories, neglecting any effect of fermionic statistics or quantum interference. 
In practice, the algorithm is a discrete-time simulation, where time is divided into small slices of equal length $\dtstep$, and the state of the system is updated only after each step, according to the events occurring in that time slice. \\
In particular, in each step the probability that a given Li atom undergoes
a collision is estimated from the local elastic scattering rate 
$\gamma(x,y,z,k) = n(x,y,z) \sigma(k) v$.  
Here $n(x,y,z)$ is the local density of the chromium gas at the position of the Li atom, $k$ is the collision wavenumber, and $v=\hbar k/\mred$ (with $\mred$ the reduced mass) denotes the relative velocity, in calculating which the Cr velocity is sampled from the thermal distribution. 
In the simulation, the dimensionless product $\Pcoll = \gamma(x,y,z,k)\dtstep$ quantifies the collision probability for a Li atom in the time interval $\dtstep$. To avoid underestimation of the number of collisions, the condition $\Pcoll \le 1$ must always be satisfied, constraining the choice of $\dtstep$. At each step, the velocity of Li is then changed with probability $\Pcoll$. The new velocity is obtained by conserving the magnitude of $v$ while changing its orientation with uniform probability across the solid angle.

The program accepts as input the main parameters characterizing the initial atomic samples, such as atom numbers, axial and radial \emph{in situ} sizes of the two clouds, trap frequencies, and effective temperatures (where the latter are intended as measurements of the variance of the corresponding velocity distributions, $\sigma_{v,i}^2 = \kBoltz\,T/m_i$, $i=$Cr, Li). Only the trajectories of Li atoms are computed, the Cr gas acting as an unperturbed thermal reservoir. 
These trajectories are used to generate histograms of the axial positions, which are fitted with Eq.~\eqref{stretchedG} to extract the mean squared displacement. Finally, the same data analysis performed on our experimental data is applied also to the simulated ones.

To understand the possible origin of the observed mismatch between simulated and experimental data in the anomalous regime, see Figs.~\ref{fig:2}c,d and \ref{fig:fig3}, we also tested the inclusion of two additional effects, on top of ISA collisions: 
(i) Wigner's delay time~\cite{Wigner1955}, which arises from the $k$-dependence of the scattering phase shift $\delta(k)$, and (ii) mean-field energy shifts, connected to the real part of the scattering amplitude~\cite{zaccanti2023}. 
The former, in the case of a free particle undergoing elastic scattering, can be written as $\Dtret = {v}^{-1} \, {\de\delta(k)}/{\de k}$, where $\cot(\delta(k)) = - (ka)^{-1} - kR^*$ for a narrow $s$-wave Feshbach resonance. 
Note that $\Dtret$ can take either positive or negative values. For harmonically trapped particles scattering off static point-like scatterers, a natural generalization of Wigner's delay time is to shift the ``undelayed'' phase-space trajectories, e.g.~in the $(\omega_x\,x\,;v_x)$ plane, by an angle $\omega_x \Dtret$ (and analogous for the $y$ and $z$ directions).
For moving scatterers, similar considerations can be applied to the motion of the relative particle. 
The mean field energy term, $\Emf(k, \,x,y,z) = -2\pi(\hbar^2/\mred) \,\Re\big[f(k)\big]\,n(x,y,z)$, is not straightforward to include in the simulator, owing to its strong $k$ dependence (here $\Re$ denotes the real part). 
A first approach is to substitute $k\rightarrow\kth$: 
In this case, $\Emf(x,y,z)$ behaves then as an additional potential energy term, with a Gaussian shape that follows from the Cr density distribution, and the equations of motion can be easily integrated numerically.
In a more refined but complicated method, the equations of motion are derived from a semiclassical Hamiltonian that includes $\Emf(k, \,x,y,z)$. 
Inclusion of Wigner's delay time and mean field energy corrections did not change qualitatively the results of the simulations.

\subsection*{Fast thermalization and role of LiCr molecules}\label{Molecules}

\begin{figure*}[t!]\centering
   \includegraphics[width=1.9 \columnwidth]{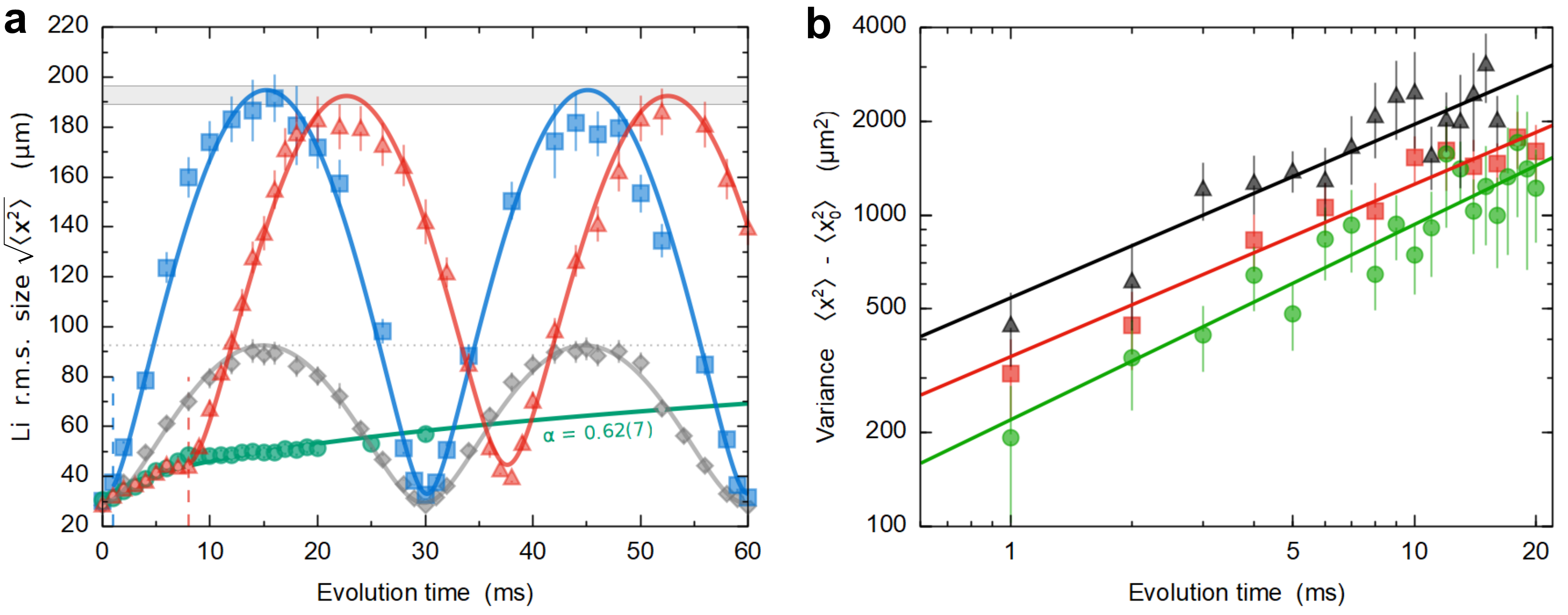}
  \caption{\textbf{Thermalization and role of molecules in the subdiffusive regime.}
   \textbf{a}, Subdiffusive data (green circles) and their best fit by Eq.~(\ref{Eq1}) (green line) are compared with the breathing dynamics of a non-interacting Li sample (gray diamonds) under identical initial conditions $T_{0,x,\text{Li}}=69(1)~\text{nK}$ and $T_{\text{Cr}}= 300(10)~\text{nK}$. 
   Blue squares (red triangles) are data taken by letting the system evolve in the interacting state for $t_{\text{int}}=1~\text{ms}$ ($t_{\text{int}}=8~\text{ms}$), after which a $0.2~\text{ms}$-long radio-frequency pulse transfers the atoms in the non-resonant Li$|2\rangle$ state. 
   Blue and red lines are fits to the data for $t>t_{\text{int}}$, see Ref.~\cite{ketterle2008}.
   The maximum reached by $\langle x^2(t) \rangle$ for $t>t_{\text{int}}$ corresponds to the value of a lithium cloud at $T_{\text{Cr}}$ (shaded area).
   \textbf{b}, Subdiffusive data in the sole infrared trap (red squares) are compared with measurements at the same $R^*/a$ value under similar Cr density and temperature conditions, but in the presence of the photo-excitation green beam (green circles), and by flipping lithium into Li$|2\rangle$ before imaging (black triangles). 
   Fits to the data (solid lines) yield power-law exponents that are compatible within fit uncertainties: 
    $\alpha=0.55(5)$ (red data), $\alpha=0.63(7)$ (green) and $\alpha=0.56(6)$ (black).} 
   \label{Extfig_FTmol}
\end{figure*}

From simple estimates, confirmed by our semi-classical simulator, in the resonant regime the Li-Cr collision rate can reach values exceeding 20~kHz, implying sub-ms thermalization times.
We confirmed this expectation in the anomalous, subdiffusive regime, by additional measurements shown in \extfig~\ref{Extfig_FTmol}a. There, we compare a standard subdiffusive trace (green dots), with the dynamics recorded by first letting lithium evolve in the interacting state for a certain time window $(0; \tint)$, after which we rapidly turned off interactions and tracked the subsequent dynamics of non-interacting Li atoms. The interaction switch is achieved by applying a 300~$\microsec$ long radio-frequency pulse, optimized to maximize the Li$|1\rangle\!\rightarrow$Li$|2\rangle$ transfer.  Blue and red dots are the results obtained for $\tint \!=$~1 and 8~ms, respectively. Gray dots, recorded on a non-interacting Li sample, serve as a reference, and their maximum value allow to determine  the $t\leq0$ effective axial temperature, $T_0$=69(1)~nK. One can notice how, instead, the  blue and red traces reach a much larger maximum value, equal for the two datasets within experimental uncertainty, yielding a gas temperature at $\tint$ of 302(4) nK that  matches the measured Cr temperature, corresponding to the shaded gray area. 
Therefore, the Li momentum distribution appears to thermalize on a sub-ms timescale, whereas the density distribution, evolving subdiffusively, is still far from reaching its equilibrium size even at our longest observation times.     

A legitimate question is whether the anomalous Li dynamics that we see at resonance may originate from molecules. For our on-resonance measurements, accepting $R^*/a\approx 0.4$ as the largest actual detuning, there are either no true molecules or molecules of size (calculated in vacuum) comparable or larger than the mean separation between Cr atoms. These estimates suggest that we are in the regime where Li propagates relatively freely by hopping from one Cr to another, as opposed to propagating as a truly bound LiCr pair. Experimentally, being unable to distinguish between these two scenarios with our absorption imaging, we addressed possible presence of molecules and their influence on the Li dynamics by using two additional protocols, applied under similar Cr temperature and density conditions in the subdiffusive regime. The first one is based on an imaging routine ``transparent'' to dimers, and the second on exploiting a green laser beam that would cause fast photo-excitation loss of dimers, if they were present.   
For the first protocol, after a variable evolution time $t$, we applied a 300~$\microsec$ long radio-frequency pulse, optimized to maximize the Li$|1\rangle\!\rightarrow$Li$|2\rangle$ atom transfer, but not able to address dimers. 
Immediately after the pulse, we imaged the \textit{in situ} distribution of transferred Li$|2\rangle$ atoms, thus tracking the $\langle x^2(t)\rangle$ dynamics of those Li particles that certainly were not in a LiCr molecular state.    
For the second protocol, we tracked the evolution of the Li$|1\rangle$ cloud in the presence of a 532 nm  laser beam, collinear with the main ODT~\cite{Ciamei2022}, set at a power level that would cause photo-excitation losses of LiCr molecules at a 200~Hz rate~\cite{Finelli2024}. 
\extfig~\ref{Extfig_FTmol}b shows the $\langle x^2(t)\rangle$ data, recorded under similar Cr gas conditions through our standard procedure (red dots), and by the two protocols just discussed, see green and black dots, respectively. A fit of Eq.~(\ref{Eq1}) to the three datasets yields $\alpha$ and $D_{\alpha}$ values identical within the fit uncertainty, see \extfig~\ref{Extfig_FTmol}b caption. 
These additional checks, together with the general considerations given above, allow us to  conclude that LiCr molecule formation cannot explain the observed anomalous dynamics.

\subsection*{Random Lorentz gas model}\label{Lorentz}

To analyze the dynamics of an atom interacting with point-like scatterers via a narrow Feshbach resonance beyond ISA, we use the standard zero-range two-channel model, which assumes that the atom can be in a closed-channel state localized on scatterer $i$ with amplitude $\phi_i$ or in a free-propagating state described by the open-channel wavefunction $\Psi({\bf r})$. The coupled Schr\"odinger equations for this problem read (for derivation see, for instance, \cite{journeaux2025})
\begin{align}
i\partial_t\Psi({\bf r},t)&=-(\nabla^2_{\bf r}/2)\Psi({\bf r},t)-	\sqrt{\pi/R^*}\sum_{i=1}^N\phi_i(t)\delta({\bf r}-{\bf R}_i),
\label{TwoBodyPsiInt}\\
i\partial_t\phi_i(t)&=-(2R^* a)^{-1}\phi_i(t)-\sqrt{\pi/R^*}\Psi_{\rm reg}({\bf R}_i,t),
\label{TwoBodyphiInt}
\end{align}
where ${\bf R}_i$ denotes the (fixed) position of the $i$-th scatterer, and the regular part of the open-channel wavefunction near this scatterer is defined as $\Psi_{\rm reg}({\bf R}_i,t)=\lim_{{\bf r}\rightarrow {\bf R}_i}\{\Psi({\bf r},t)-[\lim_{{\bf r}\rightarrow {\bf R}_i}|{\bf r}-{\bf R}_i|\Psi({\bf r},t)]/|{\bf r}-{\bf R}_i|\}$. In this section we set $\hbar=m=1$. In Eqs.~(\ref{TwoBodyPsiInt}) and (\ref{TwoBodyphiInt}) the close-to-open coupling amplitude is given by $\sqrt{\pi/R^*}$ and the detuning equals $-(2R^*a)^{-1}$.

To solve the problem in the stationary case we set $\phi_i(t)=\phi_i e^{-i\omega t}$, $\Psi({\bf r},t)=\Psi({\bf r})e^{-i\omega t}$ and eliminate the open-channel field from Eqs.~(\ref{TwoBodyPsiInt}) and (\ref{TwoBodyphiInt}) by using the equality $(-\nabla_{\bf r}^2/2-\omega)e^{i\sqrt{2\omega}r}/r=2\pi\delta({\bf r})$. The resulting equations for the amplitudes $\phi_i$ read
\begin{equation}\label{MatEq}
(1/a+i\sqrt{2\omega}+2R^*\omega)\phi_i+\sum_{j\neq i}\phi_j e^{i\sqrt{2\omega}|{\bf R}_i-{\bf R}_j|}/|{\bf R}_i-{\bf R}_j|=0.
\end{equation}  
Multiple scattering and localization in the bulk of the scattering medium can be analyzed by diagonalizing the matrix in Eq.~(\ref{MatEq}), where, for a given energy $\omega$, the quantity $-1/a$, proportional to the resonance detuning, plays the role of the eigenvalue. 

For random uncorrelated scatterers, statistical properties of the eigenstates of Eq.~(\ref{MatEq}) have been analyzed in Refs.~\cite{Skipetrov2016PRB,Skipetrov2018}. 
In particular, it is predicted that for sufficiently low energy $\omega=k^2/2$ or high density $n$ (namely, $n \gtrsim 0.08 k^3$) the eigenvectors of Eq. (\ref{MatEq}) become spatially localized if their eigenvalues fall in a range between two mobility edges. In our case, these are two values of the magnetic field detuning, as illustrated in Fig.~\ref{fig:fig4}d by white dots. These data are obtained from Ref.~\cite{Skipetrov2018}, with additional data points calculated using the same approach.

It can be seen from Eq.~(\ref{MatEq}) that stationary properties of our model are essentially governed by two dimensionless parameters constructed from the length scales $a$, $1/n^{1/3}$, and $1/k$. The width parameter $R^*$ entering through the term $2R^*\omega$ simply shifts the eigenvalue. Therefore, statistical properties of a state characterized by $R^*$ and $1/a$ are the same as statistical properties of a state with $R_1^*$, but with $1/a$ shifted by $-2(R_1^*-R^*)\omega$. An important physical consequence of this observation is that the narrower the resonance the more the region of localized states extends to the negative-$a$ side. A similar effect takes place when we increase $n$ at fixed $R^*$. Indeed, increasing $n$ and rescaling $\omega\propto n^{2/3}$ and $a\propto n^{-1/3}$ at fixed $R^*$ effectively means increasing $R^*$ and making the resonance narrower.

\subsection*{Calculation of the diffusion constant}\label{Diffusion}

This section contains details on the calculation of the diffusion constant reported in Fig.~\ref{fig:fig4}d. The procedure is based on the following arguments.

Consider the diffusion equation~\cite{LL_PhysicalKinetics} 
\begin{equation}\label{DE}
\partial_t f(t,{\bf r}) -\nabla_{\bf r}D_1\nabla_{\bf r}f(t,{\bf r})=0
\end{equation}
with constant $D_1$ inside a sphere or radius $R$ and absorptive boundary at $R$, i.e., we require $f(t,R)=0$. Assume that we drive the system at $t=0$ with a localized source at ${\bf r}={\bf r}'$. The diffusion profile is given by the Green function  $G$ satisfying
\begin{equation}\label{EqGreenclass}
\partial_t G(t,{\bf r},{\bf r}') -D_1\nabla^2_{\bf r}G(t,{\bf r},{\bf r}')=\delta(t)\delta({\bf r}-{\bf r}'),
\end{equation}
and we are interested in the survival probability $G(t,{\bf r},{\bf r})$ averaged over the sphere  
\begin{equation}\label{Pclass}
P(t)=\int G(t,{\bf r},{\bf r})d^3r={\rm Tr}e^{D_1t\nabla^2_{\bf r}}=\sum_{l,s} (2l+1)e^{-\epsilon_{l,s} D_1t/R^2},
\end{equation}
where the energies $\epsilon_{l,s}$ are eigenvalues of the operator $-\nabla^2_{\bf r}$ inside a sphere with unit radius and with Dirichlet $f=0$ boundary condition on the surface. The corresponding radial wavefunctions are $J_{l+1/2}(\sqrt{\epsilon_{l,s}}r)$ labeled by the angular momentum $l=0,1,...$ and the radial index $s=1,2,...$. The quantity $\sqrt{\epsilon_{l,s}}$ is the $s$-th zero of the Bessel function $J_{l+1/2}$. At short times $t\ll R^2/D_1$ Eq.~(\ref{Pclass}) tends to the asymptote $P(t)=R^3/[6\sqrt{\pi}(D_1t)^{3/2}]$ and in the opposite limit the sum in Eq.~(\ref{Pclass}) is dominated by the contribution of the lowest eigenvalue and we have $P(t)=e^{-\pi^2 D_1t/R^2}$. 

We now go back to Eqs.~(\ref{TwoBodyPsiInt}) and (\ref{TwoBodyphiInt}) and consider random scatterers confined in a sphere of radius $R$. We drive the $i$-th scatterer by adding to Eq.~(\ref{TwoBodyphiInt}) a Gaussian pulse with central frequency $\omega_c$ and width $\tau$. For conditions of Fig.~\ref{fig:fig4}d we use $\tau$ of order 1 ms, ensuring that we excite locally still having the energy resolution of about 10~nK. We solve the problem in the frequency domain [which amounts to solving Eq.~(\ref{MatEq}) with a Gaussian function of $\omega$ on the right-hand side] and Fourier transform back to reconstruct the trajectory $\phi_i(t)$. The survival probability $|\phi_i(t)|^2$ is then summed over $i$ and fitted with the classical curve $const\times P(t)$, the diffusion constant $D_1(\omega_c)$ emerging from fitting. Calculations are performed for $N\approx 5000$ scatterers confined in a sphere of radius $\approx 11~\micron$ and for each $\omega_c$ we average over 5 disorder realizations. We neglect finite-size effects as $R$ is much larger that the mean free path, which is of order 1 $\micron$ everywhere in Fig.~\ref{fig:fig4}d. The fitting is performed in the interval of $t$ from a few to a few tens of milliseconds. Judging by the quality of the fits we claim that the procedure is reliable in our case for $D_1\gtrsim 0.05~\hbar/m$. Approaching the localized region, the time evolution of $|\phi_i(t)|^2$ in this time interval becomes manifestly subdiffusive, and we can no longer determine $D_1$ by this method.

The diffusion constant shown in Fig.~\ref{fig:fig4}d is calculated for $n=10^{12}$~cm$^{-3}$. However, upon rescaling, this result can be used for another density of scatterers. Indeed, scaling properties of our model imply that the quantity $mD_1(R^*/a,\omega;n)/\hbar$ is a dimensionless function of only two parameters: $(1/a+2mR^*\omega/\hbar)n^{-1/3}$ and $2m\omega n^{-2/3}/\hbar$. Therefore, upon computing $D_1(R^*/a,\omega;n)$ we obtain the diffusion constant for another density $n'=n/\lambda^3$ from the relation
\begin{equation}\label{DtoD}
D_1(R^*/a,\omega;n')=D_1[\lambda R^*/a+2\lambda (1-\lambda)mR^{*2}\omega/\hbar,\lambda^2\omega;n],
\end{equation} 
which we use in our analysis when calculating $\Floceff$ and performing density averaging over the radial profile.

\newpage
\section*{Extended data}\label{secA1}
\renewcommand{\figurename}{\ExtDataFig}

\newpage

\clearpage


\begin{thebibliography}{58}%
\makeatletter
\providecommand \@ifxundefined [1]{%
 \@ifx{#1\undefined}
}%
\providecommand \@ifnum [1]{%
 \ifnum #1\expandafter \@firstoftwo
 \else \expandafter \@secondoftwo
 \fi
}%
\providecommand \@ifx [1]{%
 \ifx #1\expandafter \@firstoftwo
 \else \expandafter \@secondoftwo
 \fi
}%
\providecommand \natexlab [1]{#1}%
\providecommand \enquote  [1]{``#1''}%
\providecommand \bibnamefont  [1]{#1}%
\providecommand \bibfnamefont [1]{#1}%
\providecommand \citenamefont [1]{#1}%
\providecommand \href@noop [0]{\@secondoftwo}%
\providecommand \href [0]{\begingroup \@sanitize@url \@href}%
\providecommand \@href[1]{\@@startlink{#1}\@@href}%
\providecommand \@@href[1]{\endgroup#1\@@endlink}%
\providecommand \@sanitize@url [0]{\catcode `\\12\catcode `\$12\catcode `\&12\catcode `\#12\catcode `\^12\catcode `\_12\catcode `\%12\relax}%
\providecommand \@@startlink[1]{}%
\providecommand \@@endlink[0]{}%
\providecommand \url  [0]{\begingroup\@sanitize@url \@url }%
\providecommand \@url [1]{\endgroup\@href {#1}{\urlprefix }}%
\providecommand \urlprefix  [0]{URL }%
\providecommand \Eprint [0]{\href }%
\providecommand \doibase [0]{https://doi.org/}%
\providecommand \selectlanguage [0]{\@gobble}%
\providecommand \bibinfo  [0]{\@secondoftwo}%
\providecommand \bibfield  [0]{\@secondoftwo}%
\providecommand \translation [1]{[#1]}%
\providecommand \BibitemOpen [0]{}%
\providecommand \bibitemStop [0]{}%
\providecommand \bibitemNoStop [0]{.\EOS\space}%
\providecommand \EOS [0]{\spacefactor3000\relax}%
\providecommand \BibitemShut  [1]{\csname bibitem#1\endcsname}%
\let\auto@bib@innerbib\@empty
\bibitem [{\citenamefont {Boltzmann}(1896)}]{boltzmann}%
  \BibitemOpen
  \bibfield  {author} {\bibinfo {author} {\bibfnamefont {L.}~\bibnamefont {Boltzmann}},\ }\href@noop {} {\emph {\bibinfo {title} {Lectures on Gas Theory}}}\ (\bibinfo  {publisher} {University of California Press},\ \bibinfo {year} {1896})\BibitemShut {NoStop}%
\bibitem [{\citenamefont {Einstein}(1905)}]{einstein1905}%
  \BibitemOpen
  \bibfield  {author} {\bibinfo {author} {\bibfnamefont {A.}~\bibnamefont {Einstein}},\ }\bibfield  {title} {\bibinfo {title} {On the movement of small particles suspended in stationary liquids required by the molecular-kinetic theory of heat},\ }\href@noop {} {\bibfield  {journal} {\bibinfo  {journal} {Ann. Phys.}\ }\textbf {\bibinfo {volume} {17}},\ \bibinfo {pages} {549} (\bibinfo {year} {1905})}\BibitemShut {NoStop}%
\bibitem [{\citenamefont {Landau}\ and\ \citenamefont {Lifshitz}(1981)}]{LL_PhysicalKinetics}%
  \BibitemOpen
  \bibfield  {author} {\bibinfo {author} {\bibfnamefont {L.~D.}\ \bibnamefont {Landau}}\ and\ \bibinfo {author} {\bibfnamefont {E.~M.}\ \bibnamefont {Lifshitz}},\ }\href@noop {} {\emph {\bibinfo {title} {Physical Kinetics}}},\ \bibinfo {series} {Course of Theoretical Physics}, Vol.~\bibinfo {volume} {10}\ (\bibinfo  {publisher} {Butterworth--Heinemann},\ \bibinfo {address} {Oxford},\ \bibinfo {year} {1981})\BibitemShut {NoStop}%
\bibitem [{\citenamefont {Zwanzig}(2001)}]{zwanzig2001}%
  \BibitemOpen
  \bibfield  {author} {\bibinfo {author} {\bibfnamefont {R.}~\bibnamefont {Zwanzig}},\ }\href@noop {} {\emph {\bibinfo {title} {Nonequilibrium Statistical Mechanics}}}\ (\bibinfo  {publisher} {Oxford University Press},\ \bibinfo {address} {Oxford},\ \bibinfo {year} {2001})\BibitemShut {NoStop}%
\bibitem [{\citenamefont {Newman}(2010)}]{newman2010}%
  \BibitemOpen
  \bibfield  {author} {\bibinfo {author} {\bibfnamefont {M.~E.~J.}\ \bibnamefont {Newman}},\ }\href@noop {} {\emph {\bibinfo {title} {Networks: An Introduction}}}\ (\bibinfo  {publisher} {Oxford University Press},\ \bibinfo {address} {Oxford},\ \bibinfo {year} {2010})\BibitemShut {NoStop}%
\bibitem [{\citenamefont {Metzler}\ and\ \citenamefont {Klafter}(2000)}]{Metzler2000}%
  \BibitemOpen
  \bibfield  {author} {\bibinfo {author} {\bibfnamefont {R.}~\bibnamefont {Metzler}}\ and\ \bibinfo {author} {\bibfnamefont {J.}~\bibnamefont {Klafter}},\ }\bibfield  {title} {\bibinfo {title} {The random walk's guide to anomalous diffusion: a fractional dynamics approach},\ }\href {https://doi.org/https://doi.org/10.1016/S0370-1573(00)00070-3} {\bibfield  {journal} {\bibinfo  {journal} {Physics Reports}\ }\textbf {\bibinfo {volume} {339}},\ \bibinfo {pages} {1} (\bibinfo {year} {2000})}\BibitemShut {NoStop}%
\bibitem [{\citenamefont {Anderson}(1958)}]{anderson1958}%
  \BibitemOpen
  \bibfield  {author} {\bibinfo {author} {\bibfnamefont {P.~W.}\ \bibnamefont {Anderson}},\ }\bibfield  {title} {\bibinfo {title} {Absence of diffusion in certain random lattices},\ }\href@noop {} {\bibfield  {journal} {\bibinfo  {journal} {Phys. Rev.}\ }\textbf {\bibinfo {volume} {109}},\ \bibinfo {pages} {1492} (\bibinfo {year} {1958})}\BibitemShut {NoStop}%
\bibitem [{\citenamefont {Havlin}\ and\ \citenamefont {Ben-Avraham}(1987)}]{havlin1987}%
  \BibitemOpen
  \bibfield  {author} {\bibinfo {author} {\bibfnamefont {S.}~\bibnamefont {Havlin}}\ and\ \bibinfo {author} {\bibfnamefont {D.}~\bibnamefont {Ben-Avraham}},\ }\bibfield  {title} {\bibinfo {title} {Diffusion in disordered media},\ }\href@noop {} {\bibfield  {journal} {\bibinfo  {journal} {Adv. Phys.}\ }\textbf {\bibinfo {volume} {36}},\ \bibinfo {pages} {695} (\bibinfo {year} {1987})}\BibitemShut {NoStop}%
\bibitem [{\citenamefont {Bruun}\ and\ \citenamefont {Smith}(2007)}]{bruun2007}%
  \BibitemOpen
  \bibfield  {author} {\bibinfo {author} {\bibfnamefont {G.~M.}\ \bibnamefont {Bruun}}\ and\ \bibinfo {author} {\bibfnamefont {H.}~\bibnamefont {Smith}},\ }\bibfield  {title} {\bibinfo {title} {Shear viscosity and damping for a {F}ermi gas in the unitarity limit},\ }\href@noop {} {\bibfield  {journal} {\bibinfo  {journal} {Phys. Rev. A}\ }\textbf {\bibinfo {volume} {75}},\ \bibinfo {pages} {043612} (\bibinfo {year} {2007})}\BibitemShut {NoStop}%
\bibitem [{\citenamefont {Sommer}\ \emph {et~al.}(2011)\citenamefont {Sommer}, \citenamefont {Ku}, \citenamefont {Roati},\ and\ \citenamefont {Zwierlein}}]{Sommer2011}%
  \BibitemOpen
  \bibfield  {author} {\bibinfo {author} {\bibfnamefont {A.~T.}\ \bibnamefont {Sommer}}, \bibinfo {author} {\bibfnamefont {M.~J.~H.}\ \bibnamefont {Ku}}, \bibinfo {author} {\bibfnamefont {G.}~\bibnamefont {Roati}},\ and\ \bibinfo {author} {\bibfnamefont {M.~W.}\ \bibnamefont {Zwierlein}},\ }\bibfield  {title} {\bibinfo {title} {Universal spin transport in a strongly interacting {F}ermi gas},\ }\href {https://doi.org/10.1038/nature09989} {\bibfield  {journal} {\bibinfo  {journal} {Nature}\ }\textbf {\bibinfo {volume} {472}},\ \bibinfo {pages} {201} (\bibinfo {year} {2011})}\BibitemShut {NoStop}%
\bibitem [{\citenamefont {Landau}(1957)}]{landau1957}%
  \BibitemOpen
  \bibfield  {author} {\bibinfo {author} {\bibfnamefont {L.~D.}\ \bibnamefont {Landau}},\ }\bibfield  {title} {\bibinfo {title} {The theory of a {F}ermi liquid},\ }\href@noop {} {\bibfield  {journal} {\bibinfo  {journal} {Sov. Phys. JETP}\ }\textbf {\bibinfo {volume} {3}},\ \bibinfo {pages} {920} (\bibinfo {year} {1957})}\BibitemShut {NoStop}%
\bibitem [{\citenamefont {Baym}\ and\ \citenamefont {Pethick}(1991)}]{baym1991}%
  \BibitemOpen
  \bibfield  {author} {\bibinfo {author} {\bibfnamefont {G.}~\bibnamefont {Baym}}\ and\ \bibinfo {author} {\bibfnamefont {C.~J.}\ \bibnamefont {Pethick}},\ }\href@noop {} {\emph {\bibinfo {title} {Landau Fermi-Liquid Theory: Concepts and Applications}}}\ (\bibinfo  {publisher} {Wiley-VCH},\ \bibinfo {address} {Weinheim},\ \bibinfo {year} {1991})\BibitemShut {NoStop}%
\bibitem [{\citenamefont {Enss}\ and\ \citenamefont {Thywissen}(2019)}]{Enss2019}%
  \BibitemOpen
  \bibfield  {author} {\bibinfo {author} {\bibfnamefont {T.}~\bibnamefont {Enss}}\ and\ \bibinfo {author} {\bibfnamefont {J.~H.}\ \bibnamefont {Thywissen}},\ }\bibfield  {title} {\bibinfo {title} {Universal spin transport and quantum bounds for unitary fermions},\ }\href@noop {} {\bibfield  {journal} {\bibinfo  {journal} {Annual Review of Condensed Matter Physics}\ }\textbf {\bibinfo {volume} {10}},\ \bibinfo {pages} {85} (\bibinfo {year} {2019})}\BibitemShut {NoStop}%
\bibitem [{\citenamefont {Lorentz}(1905)}]{lorentz1905}%
  \BibitemOpen
  \bibfield  {author} {\bibinfo {author} {\bibfnamefont {H.~A.}\ \bibnamefont {Lorentz}},\ }\bibfield  {title} {\bibinfo {title} {Le mouvement des {\'e}lectrons dans les m{\'e}taux},\ }\href@noop {} {\bibfield  {journal} {\bibinfo  {journal} {Archives N\'eerlandaises des Sciences Exactes et Naturelles}\ }\textbf {\bibinfo {volume} {10}},\ \bibinfo {pages} {336} (\bibinfo {year} {1905})}\BibitemShut {NoStop}%
\bibitem [{\citenamefont {Rusek}\ \emph {et~al.}(2000)\citenamefont {Rusek}, \citenamefont {Mostowski},\ and\ \citenamefont {Or\l{}owski}}]{Rusek2000}%
  \BibitemOpen
  \bibfield  {author} {\bibinfo {author} {\bibfnamefont {M.}~\bibnamefont {Rusek}}, \bibinfo {author} {\bibfnamefont {J.}~\bibnamefont {Mostowski}},\ and\ \bibinfo {author} {\bibfnamefont {A.}~\bibnamefont {Or\l{}owski}},\ }\bibfield  {title} {\bibinfo {title} {Random {G}reen matrices: From proximity resonances to {A}nderson localization},\ }\href {https://doi.org/10.1103/PhysRevA.61.022704} {\bibfield  {journal} {\bibinfo  {journal} {Phys. Rev. A}\ }\textbf {\bibinfo {volume} {61}},\ \bibinfo {pages} {022704} (\bibinfo {year} {2000})}\BibitemShut {NoStop}%
\bibitem [{\citenamefont {Skipetrov}\ and\ \citenamefont {Sokolov}(2018)}]{Skipetrov2018}%
  \BibitemOpen
  \bibfield  {author} {\bibinfo {author} {\bibfnamefont {S.~E.}\ \bibnamefont {Skipetrov}}\ and\ \bibinfo {author} {\bibfnamefont {I.~M.}\ \bibnamefont {Sokolov}},\ }\bibfield  {title} {\bibinfo {title} {{I}offe-{R}egel criterion for {A}nderson localization in the model of resonant point scatterers},\ }\href {https://doi.org/10.1103/PhysRevB.98.064207} {\bibfield  {journal} {\bibinfo  {journal} {Phys. Rev. B}\ }\textbf {\bibinfo {volume} {98}},\ \bibinfo {pages} {064207} (\bibinfo {year} {2018})}\BibitemShut {NoStop}%
\bibitem [{\citenamefont {Smith}\ \emph {et~al.}(2017)\citenamefont {Smith}, \citenamefont {Knolle}, \citenamefont {Kovrizhin},\ and\ \citenamefont {Moessner}}]{Smith2017}%
  \BibitemOpen
  \bibfield  {author} {\bibinfo {author} {\bibfnamefont {A.}~\bibnamefont {Smith}}, \bibinfo {author} {\bibfnamefont {J.}~\bibnamefont {Knolle}}, \bibinfo {author} {\bibfnamefont {D.~L.}\ \bibnamefont {Kovrizhin}},\ and\ \bibinfo {author} {\bibfnamefont {R.}~\bibnamefont {Moessner}},\ }\bibfield  {title} {\bibinfo {title} {Disorder-free localization},\ }\href {https://doi.org/10.1103/PhysRevLett.118.266601} {\bibfield  {journal} {\bibinfo  {journal} {Phys. Rev. Lett.}\ }\textbf {\bibinfo {volume} {118}},\ \bibinfo {pages} {266601} (\bibinfo {year} {2017})}\BibitemShut {NoStop}%
\bibitem [{\citenamefont {Grover}\ and\ \citenamefont {Fisher}(2014)}]{Grover2014}%
  \BibitemOpen
  \bibfield  {author} {\bibinfo {author} {\bibfnamefont {T.}~\bibnamefont {Grover}}\ and\ \bibinfo {author} {\bibfnamefont {M.~P.~A.}\ \bibnamefont {Fisher}},\ }\bibfield  {title} {\bibinfo {title} {Quantum disentangled liquids},\ }\href {https://doi.org/10.1088/1742-5468/2014/10/P10010} {\bibfield  {journal} {\bibinfo  {journal} {Journal of Statistical Mechanics: Theory and Experiment}\ }\textbf {\bibinfo {volume} {2014}},\ \bibinfo {pages} {P10010} (\bibinfo {year} {2014})}\BibitemShut {NoStop}%
\bibitem [{\citenamefont {Ashcroft}\ and\ \citenamefont {Mermin}(1976)}]{Ashcroft76}%
  \BibitemOpen
  \bibfield  {author} {\bibinfo {author} {\bibfnamefont {N.~W.}\ \bibnamefont {Ashcroft}}\ and\ \bibinfo {author} {\bibfnamefont {N.~D.}\ \bibnamefont {Mermin}},\ }\href@noop {} {\emph {\bibinfo {title} {{S}olid {S}tate {P}hysics}}}\ (\bibinfo  {publisher} {Holt-Saunders},\ \bibinfo {year} {1976})\BibitemShut {NoStop}%
\bibitem [{\citenamefont {Zwerger}(2012)}]{zwerger}%
  \BibitemOpen
  \bibinfo {editor} {\bibfnamefont {W.}~\bibnamefont {Zwerger}},\ ed.,\ \href@noop {} {\emph {\bibinfo {title} {The BCS-BEC Crossover and the Unitary {F}ermi Gas}}}\ (\bibinfo  {publisher} {Springer},\ \bibinfo {year} {2012})\BibitemShut {NoStop}%
\bibitem [{\citenamefont {Mott}(1990)}]{mott1990}%
  \BibitemOpen
  \bibfield  {author} {\bibinfo {author} {\bibfnamefont {N.}~\bibnamefont {Mott}},\ }\href@noop {} {\emph {\bibinfo {title} {Metal-Insulator Transitions}}}\ (\bibinfo  {publisher} {Taylor \& Francis},\ \bibinfo {year} {1990})\BibitemShut {NoStop}%
\bibitem [{\citenamefont {Dean}\ \emph {et~al.}(2013)\citenamefont {Dean}, \citenamefont {Wang}, \citenamefont {Maher}, \citenamefont {Forsythe}, \citenamefont {Ghahari}, \citenamefont {Gao}, \citenamefont {Katoch}, \citenamefont {Ishigami}, \citenamefont {Moon}, \citenamefont {Koshino}, \citenamefont {Taniguchi}, \citenamefont {Watanabe}, \citenamefont {Shepard}, \citenamefont {Hone},\ and\ \citenamefont {Kim}}]{dean}%
  \BibitemOpen
  \bibfield  {author} {\bibinfo {author} {\bibfnamefont {C.~R.}\ \bibnamefont {Dean}}, \bibinfo {author} {\bibfnamefont {L.}~\bibnamefont {Wang}}, \bibinfo {author} {\bibfnamefont {P.}~\bibnamefont {Maher}}, \bibinfo {author} {\bibfnamefont {C.}~\bibnamefont {Forsythe}}, \bibinfo {author} {\bibfnamefont {F.}~\bibnamefont {Ghahari}}, \bibinfo {author} {\bibfnamefont {Y.}~\bibnamefont {Gao}}, \bibinfo {author} {\bibfnamefont {J.}~\bibnamefont {Katoch}}, \bibinfo {author} {\bibfnamefont {M.}~\bibnamefont {Ishigami}}, \bibinfo {author} {\bibfnamefont {P.}~\bibnamefont {Moon}}, \bibinfo {author} {\bibfnamefont {M.}~\bibnamefont {Koshino}}, \bibinfo {author} {\bibfnamefont {T.}~\bibnamefont {Taniguchi}}, \bibinfo {author} {\bibfnamefont {K.}~\bibnamefont {Watanabe}}, \bibinfo {author} {\bibfnamefont {K.~L.}\ \bibnamefont {Shepard}}, \bibinfo {author} {\bibfnamefont {J.}~\bibnamefont {Hone}},\ and\ \bibinfo {author} {\bibfnamefont {P.}~\bibnamefont {Kim}},\ }\bibfield  {title} {\bibinfo {title} {Hofstadter’s
  butterfly and the fractal quantum hall effect in moiré superlattices},\ }\href@noop {} {\bibfield  {journal} {\bibinfo  {journal} {Nature}\ }\textbf {\bibinfo {volume} {497}},\ \bibinfo {pages} {598} (\bibinfo {year} {2013})}\BibitemShut {NoStop}%
\bibitem [{\citenamefont {Ciamei}\ \emph {et~al.}(2022{\natexlab{a}})\citenamefont {Ciamei}, \citenamefont {Finelli}, \citenamefont {Cosco}, \citenamefont {Inguscio}, \citenamefont {Trenkwalder},\ and\ \citenamefont {Zaccanti}}]{Ciamei2022}%
  \BibitemOpen
  \bibfield  {author} {\bibinfo {author} {\bibfnamefont {A.}~\bibnamefont {Ciamei}}, \bibinfo {author} {\bibfnamefont {S.}~\bibnamefont {Finelli}}, \bibinfo {author} {\bibfnamefont {A.}~\bibnamefont {Cosco}}, \bibinfo {author} {\bibfnamefont {M.}~\bibnamefont {Inguscio}}, \bibinfo {author} {\bibfnamefont {A.}~\bibnamefont {Trenkwalder}},\ and\ \bibinfo {author} {\bibfnamefont {M.}~\bibnamefont {Zaccanti}},\ }\bibfield  {title} {\bibinfo {title} {Double-degenerate {F}ermi mixtures of $^{6}\mathrm{Li}$ and $^{53}\mathrm{Cr}$ atoms},\ }\href {https://doi.org/10.1103/PhysRevA.106.053318} {\bibfield  {journal} {\bibinfo  {journal} {Phys. Rev. A}\ }\textbf {\bibinfo {volume} {106}},\ \bibinfo {pages} {053318} (\bibinfo {year} {2022}{\natexlab{a}})}\BibitemShut {NoStop}%
\bibitem [{\citenamefont {Ciamei}\ \emph {et~al.}(2022{\natexlab{b}})\citenamefont {Ciamei}, \citenamefont {Finelli}, \citenamefont {Trenkwalder}, \citenamefont {Inguscio}, \citenamefont {Simoni},\ and\ \citenamefont {Zaccanti}}]{Ciamei2022A}%
  \BibitemOpen
  \bibfield  {author} {\bibinfo {author} {\bibfnamefont {A.}~\bibnamefont {Ciamei}}, \bibinfo {author} {\bibfnamefont {S.}~\bibnamefont {Finelli}}, \bibinfo {author} {\bibfnamefont {A.}~\bibnamefont {Trenkwalder}}, \bibinfo {author} {\bibfnamefont {M.}~\bibnamefont {Inguscio}}, \bibinfo {author} {\bibfnamefont {A.}~\bibnamefont {Simoni}},\ and\ \bibinfo {author} {\bibfnamefont {M.}~\bibnamefont {Zaccanti}},\ }\bibfield  {title} {\bibinfo {title} {Exploring ultracold collisions in $^{6}\mathrm{Li}\text{\ensuremath{-}}^{53}\mathrm{Cr}$ {F}ermi mixtures: {F}eshbach resonances and scattering properties of a novel alkali-transition metal system},\ }\href {https://doi.org/10.1103/PhysRevLett.129.093402} {\bibfield  {journal} {\bibinfo  {journal} {Phys. Rev. Lett.}\ }\textbf {\bibinfo {volume} {129}},\ \bibinfo {pages} {093402} (\bibinfo {year} {2022}{\natexlab{b}})}\BibitemShut {NoStop}%
\bibitem [{\citenamefont {Bruun}(2011)}]{Bruun2011}%
  \BibitemOpen
  \bibfield  {author} {\bibinfo {author} {\bibfnamefont {G.~M.}\ \bibnamefont {Bruun}},\ }\bibfield  {title} {\bibinfo {title} {Spin diffusion in {F}ermi gases},\ }\href {https://doi.org/10.1088/1367-2630/13/3/035005} {\bibfield  {journal} {\bibinfo  {journal} {New Journal of Physics}\ }\textbf {\bibinfo {volume} {13}},\ \bibinfo {pages} {035005} (\bibinfo {year} {2011})}\BibitemShut {NoStop}%
\bibitem [{\citenamefont {Goulko}\ \emph {et~al.}(2013)\citenamefont {Goulko}, \citenamefont {Chevy},\ and\ \citenamefont {Lobo}}]{Goulko2013}%
  \BibitemOpen
  \bibfield  {author} {\bibinfo {author} {\bibfnamefont {O.}~\bibnamefont {Goulko}}, \bibinfo {author} {\bibfnamefont {F.}~\bibnamefont {Chevy}},\ and\ \bibinfo {author} {\bibfnamefont {C.}~\bibnamefont {Lobo}},\ }\bibfield  {title} {\bibinfo {title} {Spin drag of a {F}ermi gas in a harmonic trap},\ }\href {https://doi.org/10.1103/PhysRevLett.111.190402} {\bibfield  {journal} {\bibinfo  {journal} {Phys. Rev. Lett.}\ }\textbf {\bibinfo {volume} {111}},\ \bibinfo {pages} {190402} (\bibinfo {year} {2013})}\BibitemShut {NoStop}%
\bibitem [{\citenamefont {Finelli}\ \emph {et~al.}(2024)\citenamefont {Finelli}, \citenamefont {Ciamei}, \citenamefont {Restivo}, \citenamefont {Schemmer}, \citenamefont {Cosco}, \citenamefont {Inguscio}, \citenamefont {Trenkwalder}, \citenamefont {Zaremba-Kopczyk}, \citenamefont {Gronowski}, \citenamefont {Tomza},\ and\ \citenamefont {Zaccanti}}]{Finelli2024}%
  \BibitemOpen
  \bibfield  {author} {\bibinfo {author} {\bibfnamefont {S.}~\bibnamefont {Finelli}}, \bibinfo {author} {\bibfnamefont {A.}~\bibnamefont {Ciamei}}, \bibinfo {author} {\bibfnamefont {B.}~\bibnamefont {Restivo}}, \bibinfo {author} {\bibfnamefont {M.}~\bibnamefont {Schemmer}}, \bibinfo {author} {\bibfnamefont {A.}~\bibnamefont {Cosco}}, \bibinfo {author} {\bibfnamefont {M.}~\bibnamefont {Inguscio}}, \bibinfo {author} {\bibfnamefont {A.}~\bibnamefont {Trenkwalder}}, \bibinfo {author} {\bibfnamefont {K.}~\bibnamefont {Zaremba-Kopczyk}}, \bibinfo {author} {\bibfnamefont {M.}~\bibnamefont {Gronowski}}, \bibinfo {author} {\bibfnamefont {M.}~\bibnamefont {Tomza}},\ and\ \bibinfo {author} {\bibfnamefont {M.}~\bibnamefont {Zaccanti}},\ }\bibfield  {title} {\bibinfo {title} {Ultracold $\mathrm{Li}\mathrm{Cr}$: A new pathway to quantum gases of paramagnetic polar molecules},\ }\href {https://doi.org/10.1103/PRXQuantum.5.020358} {\bibfield  {journal} {\bibinfo  {journal} {PRX Quantum}\ }\textbf {\bibinfo {volume} {5}},\
  \bibinfo {pages} {020358} (\bibinfo {year} {2024})}\BibitemShut {NoStop}%
\bibitem [{\citenamefont {Breit}\ and\ \citenamefont {Wigner}(1936)}]{Breit1936}%
  \BibitemOpen
  \bibfield  {author} {\bibinfo {author} {\bibfnamefont {G.}~\bibnamefont {Breit}}\ and\ \bibinfo {author} {\bibfnamefont {E.}~\bibnamefont {Wigner}},\ }\bibfield  {title} {\bibinfo {title} {Capture of slow neutrons},\ }\href {https://doi.org/10.1103/PhysRev.49.519} {\bibfield  {journal} {\bibinfo  {journal} {Phys. Rev.}\ }\textbf {\bibinfo {volume} {49}},\ \bibinfo {pages} {519} (\bibinfo {year} {1936})}\BibitemShut {NoStop}%
\bibitem [{\citenamefont {Chin}\ \emph {et~al.}(2010)\citenamefont {Chin}, \citenamefont {Grimm}, \citenamefont {Julienne},\ and\ \citenamefont {Tiesinga}}]{chin2010}%
  \BibitemOpen
  \bibfield  {author} {\bibinfo {author} {\bibfnamefont {C.}~\bibnamefont {Chin}}, \bibinfo {author} {\bibfnamefont {R.}~\bibnamefont {Grimm}}, \bibinfo {author} {\bibfnamefont {P.}~\bibnamefont {Julienne}},\ and\ \bibinfo {author} {\bibfnamefont {E.}~\bibnamefont {Tiesinga}},\ }\bibfield  {title} {\bibinfo {title} {Feshbach resonances in ultracold gases},\ }\href@noop {} {\bibfield  {journal} {\bibinfo  {journal} {Rev. Mod. Phys.}\ }\textbf {\bibinfo {volume} {82}},\ \bibinfo {pages} {1225} (\bibinfo {year} {2010})}\BibitemShut {NoStop}%
\bibitem [{\citenamefont {Uhlenbeck}\ and\ \citenamefont {Ornstein}(1930)}]{Uhlenbeck1930}%
  \BibitemOpen
  \bibfield  {author} {\bibinfo {author} {\bibfnamefont {G.~E.}\ \bibnamefont {Uhlenbeck}}\ and\ \bibinfo {author} {\bibfnamefont {L.~S.}\ \bibnamefont {Ornstein}},\ }\bibfield  {title} {\bibinfo {title} {On the theory of the {B}rownian motion},\ }\href {https://doi.org/10.1103/PhysRev.36.823} {\bibfield  {journal} {\bibinfo  {journal} {Phys. Rev.}\ }\textbf {\bibinfo {volume} {36}},\ \bibinfo {pages} {823} (\bibinfo {year} {1930})}\BibitemShut {NoStop}%
\bibitem [{\citenamefont {Barbosa}\ \emph {et~al.}(2025)\citenamefont {Barbosa}, \citenamefont {Kiefer-Emmanouilidis}, \citenamefont {Lang}, \citenamefont {Koch},\ and\ \citenamefont {Widera}}]{Barbosa2024}%
  \BibitemOpen
  \bibfield  {author} {\bibinfo {author} {\bibfnamefont {S.}~\bibnamefont {Barbosa}}, \bibinfo {author} {\bibfnamefont {M.}~\bibnamefont {Kiefer-Emmanouilidis}}, \bibinfo {author} {\bibfnamefont {F.}~\bibnamefont {Lang}}, \bibinfo {author} {\bibfnamefont {J.}~\bibnamefont {Koch}},\ and\ \bibinfo {author} {\bibfnamefont {A.}~\bibnamefont {Widera}},\ }\bibfield  {title} {\bibinfo {title} {Stabilizing an ultracold {F}ermi gas against {F}ermi acceleration to superdiffusion through localization},\ }\href {https://doi.org/10.1103/rq4w-377l} {\bibfield  {journal} {\bibinfo  {journal} {Phys. Rev. Lett.}\ }\textbf {\bibinfo {volume} {134}},\ \bibinfo {pages} {253402} (\bibinfo {year} {2025})}\BibitemShut {NoStop}%
\bibitem [{\citenamefont {Zaccanti}(2023)}]{zaccanti2023}%
  \BibitemOpen
  \bibfield  {author} {\bibinfo {author} {\bibfnamefont {M.}~\bibnamefont {Zaccanti}},\ }\bibfield  {title} {\bibinfo {title} {Mass‐imbalanced {F}ermi mixtures with resonant interactions},\ }\href@noop {} {\bibfield  {journal} {\bibinfo  {journal} {Lecture Notes of the Enrico Fermi Summer School: Quantum Mixtures with Ultra-Cold Atoms}\ } (\bibinfo {year} {2023})},\ \bibinfo {note} {arXiv:2306.02736; Lecture notes for Varenna School}\BibitemShut {NoStop}%
\bibitem [{\citenamefont {Valtolina}\ \emph {et~al.}(2017)\citenamefont {Valtolina}, \citenamefont {Scazza}, \citenamefont {Amico}, \citenamefont {Burchianti}, \citenamefont {Recati}, \citenamefont {Enss}, \citenamefont {Inguscio}, \citenamefont {Zaccanti},\ and\ \citenamefont {Roati}}]{Valtolina2017}%
  \BibitemOpen
  \bibfield  {author} {\bibinfo {author} {\bibfnamefont {G.}~\bibnamefont {Valtolina}}, \bibinfo {author} {\bibfnamefont {F.}~\bibnamefont {Scazza}}, \bibinfo {author} {\bibfnamefont {A.}~\bibnamefont {Amico}}, \bibinfo {author} {\bibfnamefont {A.}~\bibnamefont {Burchianti}}, \bibinfo {author} {\bibfnamefont {A.}~\bibnamefont {Recati}}, \bibinfo {author} {\bibfnamefont {T.}~\bibnamefont {Enss}}, \bibinfo {author} {\bibfnamefont {M.}~\bibnamefont {Inguscio}}, \bibinfo {author} {\bibfnamefont {M.}~\bibnamefont {Zaccanti}},\ and\ \bibinfo {author} {\bibfnamefont {G.}~\bibnamefont {Roati}},\ }\bibfield  {title} {\bibinfo {title} {Exploring the ferromagnetic behaviour of a repulsive {F}ermi gas through spin dynamics},\ }\href {https://doi.org/10.1038/nphys4108} {\bibfield  {journal} {\bibinfo  {journal} {Nature Physics}\ }\textbf {\bibinfo {volume} {13}},\ \bibinfo {pages} {704} (\bibinfo {year} {2017})}\BibitemShut {NoStop}%
\bibitem [{\citenamefont {Billy}\ \emph {et~al.}(2008)\citenamefont {Billy}, \citenamefont {Josse}, \citenamefont {Zuo}, \citenamefont {Bernard}, \citenamefont {Hambrecht}, \citenamefont {Lugan}, \citenamefont {Cl{\'e}ment}, \citenamefont {Sanchez-Palencia}, \citenamefont {Bouyer},\ and\ \citenamefont {Aspect}}]{Billy2008}%
  \BibitemOpen
  \bibfield  {author} {\bibinfo {author} {\bibfnamefont {J.}~\bibnamefont {Billy}}, \bibinfo {author} {\bibfnamefont {V.}~\bibnamefont {Josse}}, \bibinfo {author} {\bibfnamefont {Z.}~\bibnamefont {Zuo}}, \bibinfo {author} {\bibfnamefont {A.}~\bibnamefont {Bernard}}, \bibinfo {author} {\bibfnamefont {B.}~\bibnamefont {Hambrecht}}, \bibinfo {author} {\bibfnamefont {P.}~\bibnamefont {Lugan}}, \bibinfo {author} {\bibfnamefont {D.}~\bibnamefont {Cl{\'e}ment}}, \bibinfo {author} {\bibfnamefont {L.}~\bibnamefont {Sanchez-Palencia}}, \bibinfo {author} {\bibfnamefont {P.}~\bibnamefont {Bouyer}},\ and\ \bibinfo {author} {\bibfnamefont {A.}~\bibnamefont {Aspect}},\ }\bibfield  {title} {\bibinfo {title} {Direct observation of {A}nderson localization of matter waves in a controlled disorder},\ }\href {https://doi.org/10.1038/nature07000} {\bibfield  {journal} {\bibinfo  {journal} {Nature}\ }\textbf {\bibinfo {volume} {453}},\ \bibinfo {pages} {891} (\bibinfo {year} {2008})}\BibitemShut {NoStop}%
\bibitem [{\citenamefont {Roati}\ \emph {et~al.}(2008)\citenamefont {Roati}, \citenamefont {D'Errico}, \citenamefont {Fallani}, \citenamefont {Fattori}, \citenamefont {Fort}, \citenamefont {Zaccanti}, \citenamefont {Modugno}, \citenamefont {Modugno},\ and\ \citenamefont {Inguscio}}]{Roati2008}%
  \BibitemOpen
  \bibfield  {author} {\bibinfo {author} {\bibfnamefont {G.}~\bibnamefont {Roati}}, \bibinfo {author} {\bibfnamefont {C.}~\bibnamefont {D'Errico}}, \bibinfo {author} {\bibfnamefont {L.}~\bibnamefont {Fallani}}, \bibinfo {author} {\bibfnamefont {M.}~\bibnamefont {Fattori}}, \bibinfo {author} {\bibfnamefont {C.}~\bibnamefont {Fort}}, \bibinfo {author} {\bibfnamefont {M.}~\bibnamefont {Zaccanti}}, \bibinfo {author} {\bibfnamefont {G.}~\bibnamefont {Modugno}}, \bibinfo {author} {\bibfnamefont {M.}~\bibnamefont {Modugno}},\ and\ \bibinfo {author} {\bibfnamefont {M.}~\bibnamefont {Inguscio}},\ }\bibfield  {title} {\bibinfo {title} {{A}nderson localization of a non-interacting {B}ose--{E}instein condensate},\ }\href {https://doi.org/10.1038/nature07071} {\bibfield  {journal} {\bibinfo  {journal} {Nature}\ }\textbf {\bibinfo {volume} {453}},\ \bibinfo {pages} {895} (\bibinfo {year} {2008})}\BibitemShut {NoStop}%
\bibitem [{\citenamefont {Chab{\'e}}\ \emph {et~al.}(2008)\citenamefont {Chab{\'e}}, \citenamefont {Lemari{\'e}}, \citenamefont {Gr{\'e}maud}, \citenamefont {Delande}, \citenamefont {Szriftgiser},\ and\ \citenamefont {Garreau}}]{Chabe2008}%
  \BibitemOpen
  \bibfield  {author} {\bibinfo {author} {\bibfnamefont {J.}~\bibnamefont {Chab{\'e}}}, \bibinfo {author} {\bibfnamefont {G.}~\bibnamefont {Lemari{\'e}}}, \bibinfo {author} {\bibfnamefont {B.}~\bibnamefont {Gr{\'e}maud}}, \bibinfo {author} {\bibfnamefont {D.}~\bibnamefont {Delande}}, \bibinfo {author} {\bibfnamefont {P.}~\bibnamefont {Szriftgiser}},\ and\ \bibinfo {author} {\bibfnamefont {J.~C.}\ \bibnamefont {Garreau}},\ }\bibfield  {title} {\bibinfo {title} {Experimental observation of the {A}nderson metal-insulator transition with atomic matter waves},\ }\href@noop {} {\bibfield  {journal} {\bibinfo  {journal} {Phys. Rev. Lett.}\ }\textbf {\bibinfo {volume} {101}},\ \bibinfo {pages} {255702} (\bibinfo {year} {2008})}\BibitemShut {NoStop}%
\bibitem [{\citenamefont {Hu}\ \emph {et~al.}(2008)\citenamefont {Hu}, \citenamefont {Strybulevych}, \citenamefont {Page}, \citenamefont {Skipetrov},\ and\ \citenamefont {van Tiggelen}}]{Hu2008}%
  \BibitemOpen
  \bibfield  {author} {\bibinfo {author} {\bibfnamefont {H.}~\bibnamefont {Hu}}, \bibinfo {author} {\bibfnamefont {A.}~\bibnamefont {Strybulevych}}, \bibinfo {author} {\bibfnamefont {J.~H.}\ \bibnamefont {Page}}, \bibinfo {author} {\bibfnamefont {S.~E.}\ \bibnamefont {Skipetrov}},\ and\ \bibinfo {author} {\bibfnamefont {B.~A.}\ \bibnamefont {van Tiggelen}},\ }\bibfield  {title} {\bibinfo {title} {Localization of ultrasound in a three-dimensional elastic network},\ }\href {https://doi.org/10.1038/nphys1101} {\bibfield  {journal} {\bibinfo  {journal} {Nature Physics}\ }\textbf {\bibinfo {volume} {4}},\ \bibinfo {pages} {945} (\bibinfo {year} {2008})}\BibitemShut {NoStop}%
\bibitem [{\citenamefont {Semeghini}\ \emph {et~al.}(2015)\citenamefont {Semeghini}, \citenamefont {Landini}, \citenamefont {Castilho}, \citenamefont {Roy}, \citenamefont {Spagnolli}, \citenamefont {Trenkwalder}, \citenamefont {Fattori}, \citenamefont {Inguscio},\ and\ \citenamefont {Modugno}}]{Semeghini2015}%
  \BibitemOpen
  \bibfield  {author} {\bibinfo {author} {\bibfnamefont {G.}~\bibnamefont {Semeghini}}, \bibinfo {author} {\bibfnamefont {M.}~\bibnamefont {Landini}}, \bibinfo {author} {\bibfnamefont {P.}~\bibnamefont {Castilho}}, \bibinfo {author} {\bibfnamefont {S.}~\bibnamefont {Roy}}, \bibinfo {author} {\bibfnamefont {G.}~\bibnamefont {Spagnolli}}, \bibinfo {author} {\bibfnamefont {A.}~\bibnamefont {Trenkwalder}}, \bibinfo {author} {\bibfnamefont {M.}~\bibnamefont {Fattori}}, \bibinfo {author} {\bibfnamefont {M.}~\bibnamefont {Inguscio}},\ and\ \bibinfo {author} {\bibfnamefont {G.}~\bibnamefont {Modugno}},\ }\bibfield  {title} {\bibinfo {title} {Measurement of the mobility edge for 3{D} {A}nderson localization},\ }\href {https://doi.org/10.1038/nphys3339} {\bibfield  {journal} {\bibinfo  {journal} {Nature Physics}\ }\textbf {\bibinfo {volume} {11}},\ \bibinfo {pages} {554} (\bibinfo {year} {2015})}\BibitemShut {NoStop}%
\bibitem [{\citenamefont {Cobus}\ \emph {et~al.}(2018)\citenamefont {Cobus}, \citenamefont {Hildebrand}, \citenamefont {Skipetrov}, \citenamefont {van Tiggelen},\ and\ \citenamefont {Page}}]{Cobus2018}%
  \BibitemOpen
  \bibfield  {author} {\bibinfo {author} {\bibfnamefont {L.~A.}\ \bibnamefont {Cobus}}, \bibinfo {author} {\bibfnamefont {W.~K.}\ \bibnamefont {Hildebrand}}, \bibinfo {author} {\bibfnamefont {S.~E.}\ \bibnamefont {Skipetrov}}, \bibinfo {author} {\bibfnamefont {B.~A.}\ \bibnamefont {van Tiggelen}},\ and\ \bibinfo {author} {\bibfnamefont {J.~H.}\ \bibnamefont {Page}},\ }\bibfield  {title} {\bibinfo {title} {Transverse confinement of ultrasound through the {A}nderson transition in three-dimensional mesoglasses},\ }\href {https://doi.org/10.1103/PhysRevB.98.214201} {\bibfield  {journal} {\bibinfo  {journal} {Phys. Rev. B}\ }\textbf {\bibinfo {volume} {98}},\ \bibinfo {pages} {214201} (\bibinfo {year} {2018})}\BibitemShut {NoStop}%
\bibitem [{\citenamefont {Darkwah~Oppong}\ \emph {et~al.}(2022)\citenamefont {Darkwah~Oppong}, \citenamefont {Pasqualetti}, \citenamefont {Bettermann}, \citenamefont {Zechmann}, \citenamefont {Knap}, \citenamefont {Bloch},\ and\ \citenamefont {F\"olling}}]{Oppong2022}%
  \BibitemOpen
  \bibfield  {author} {\bibinfo {author} {\bibfnamefont {N.}~\bibnamefont {Darkwah~Oppong}}, \bibinfo {author} {\bibfnamefont {G.}~\bibnamefont {Pasqualetti}}, \bibinfo {author} {\bibfnamefont {O.}~\bibnamefont {Bettermann}}, \bibinfo {author} {\bibfnamefont {P.}~\bibnamefont {Zechmann}}, \bibinfo {author} {\bibfnamefont {M.}~\bibnamefont {Knap}}, \bibinfo {author} {\bibfnamefont {I.}~\bibnamefont {Bloch}},\ and\ \bibinfo {author} {\bibfnamefont {S.}~\bibnamefont {F\"olling}},\ }\bibfield  {title} {\bibinfo {title} {Probing transport and slow relaxation in the mass-imbalanced {F}ermi-{H}ubbard model},\ }\href {https://doi.org/10.1103/PhysRevX.12.031026} {\bibfield  {journal} {\bibinfo  {journal} {Phys. Rev. X}\ }\textbf {\bibinfo {volume} {12}},\ \bibinfo {pages} {031026} (\bibinfo {year} {2022})}\BibitemShut {NoStop}%
\bibitem [{\citenamefont {Lee}\ and\ \citenamefont {Ramakrishnan}(1985)}]{Lee1985}%
  \BibitemOpen
  \bibfield  {author} {\bibinfo {author} {\bibfnamefont {P.~A.}\ \bibnamefont {Lee}}\ and\ \bibinfo {author} {\bibfnamefont {T.~V.}\ \bibnamefont {Ramakrishnan}},\ }\bibfield  {title} {\bibinfo {title} {Disordered electronic systems},\ }\href {https://doi.org/10.1103/RevModPhys.57.287} {\bibfield  {journal} {\bibinfo  {journal} {Rev. Mod. Phys.}\ }\textbf {\bibinfo {volume} {57}},\ \bibinfo {pages} {287} (\bibinfo {year} {1985})}\BibitemShut {NoStop}%
\bibitem [{\citenamefont {Kramer}\ and\ \citenamefont {MacKinnon}(1993)}]{Kramer1993}%
  \BibitemOpen
  \bibfield  {author} {\bibinfo {author} {\bibfnamefont {B.}~\bibnamefont {Kramer}}\ and\ \bibinfo {author} {\bibfnamefont {A.}~\bibnamefont {MacKinnon}},\ }\bibfield  {title} {\bibinfo {title} {Localization: theory and experiment},\ }\href {https://doi.org/10.1088/0034-4885/56/12/001} {\bibfield  {journal} {\bibinfo  {journal} {Reports on Progress in Physics}\ }\textbf {\bibinfo {volume} {56}},\ \bibinfo {pages} {1469} (\bibinfo {year} {1993})}\BibitemShut {NoStop}%
\bibitem [{\citenamefont {Basko}\ \emph {et~al.}(2006)\citenamefont {Basko}, \citenamefont {Aleiner},\ and\ \citenamefont {Altshuler}}]{Basko2006}%
  \BibitemOpen
  \bibfield  {author} {\bibinfo {author} {\bibfnamefont {D.}~\bibnamefont {Basko}}, \bibinfo {author} {\bibfnamefont {I.}~\bibnamefont {Aleiner}},\ and\ \bibinfo {author} {\bibfnamefont {B.}~\bibnamefont {Altshuler}},\ }\bibfield  {title} {\bibinfo {title} {Metal–insulator transition in a weakly interacting many-electron system with localized single-particle states},\ }\href {https://doi.org/https://doi.org/10.1016/j.aop.2005.11.014} {\bibfield  {journal} {\bibinfo  {journal} {Annals of Physics}\ }\textbf {\bibinfo {volume} {321}},\ \bibinfo {pages} {1126} (\bibinfo {year} {2006})}\BibitemShut {NoStop}%
\bibitem [{\citenamefont {Abanin}\ \emph {et~al.}(2019)\citenamefont {Abanin}, \citenamefont {Altman}, \citenamefont {Bloch},\ and\ \citenamefont {Serbyn}}]{Abanin2019}%
  \BibitemOpen
  \bibfield  {author} {\bibinfo {author} {\bibfnamefont {D.~A.}\ \bibnamefont {Abanin}}, \bibinfo {author} {\bibfnamefont {E.}~\bibnamefont {Altman}}, \bibinfo {author} {\bibfnamefont {I.}~\bibnamefont {Bloch}},\ and\ \bibinfo {author} {\bibfnamefont {M.}~\bibnamefont {Serbyn}},\ }\bibfield  {title} {\bibinfo {title} {Colloquium: Many-body localization, thermalization, and entanglement},\ }\href {https://doi.org/10.1103/RevModPhys.91.021001} {\bibfield  {journal} {\bibinfo  {journal} {Rev. Mod. Phys.}\ }\textbf {\bibinfo {volume} {91}},\ \bibinfo {pages} {021001} (\bibinfo {year} {2019})}\BibitemShut {NoStop}%
\bibitem [{\citenamefont {Massignan}\ and\ \citenamefont {Castin}(2006)}]{Massignan2006}%
  \BibitemOpen
  \bibfield  {author} {\bibinfo {author} {\bibfnamefont {P.}~\bibnamefont {Massignan}}\ and\ \bibinfo {author} {\bibfnamefont {Y.}~\bibnamefont {Castin}},\ }\bibfield  {title} {\bibinfo {title} {Three-dimensional strong localization of matter waves by scattering from atoms in a lattice with a confinement-induced resonance},\ }\href {https://doi.org/10.1103/PhysRevA.74.013616} {\bibfield  {journal} {\bibinfo  {journal} {Phys. Rev. A}\ }\textbf {\bibinfo {volume} {74}},\ \bibinfo {pages} {013616} (\bibinfo {year} {2006})}\BibitemShut {NoStop}%
\bibitem [{\citenamefont {Yao}\ \emph {et~al.}(2025)\citenamefont {Yao}, \citenamefont {Chi}, \citenamefont {Wang}, \citenamefont {Fletcher},\ and\ \citenamefont {Zwierlein}}]{Yao2025}%
  \BibitemOpen
  \bibfield  {author} {\bibinfo {author} {\bibfnamefont {R.}~\bibnamefont {Yao}}, \bibinfo {author} {\bibfnamefont {S.}~\bibnamefont {Chi}}, \bibinfo {author} {\bibfnamefont {M.}~\bibnamefont {Wang}}, \bibinfo {author} {\bibfnamefont {R.~J.}\ \bibnamefont {Fletcher}},\ and\ \bibinfo {author} {\bibfnamefont {M.}~\bibnamefont {Zwierlein}},\ }\bibfield  {title} {\bibinfo {title} {Measuring pair correlations in bose and {F}ermi gases via atom-resolved microscopy},\ }\href {https://doi.org/10.1103/PhysRevLett.134.183402} {\bibfield  {journal} {\bibinfo  {journal} {Phys. Rev. Lett.}\ }\textbf {\bibinfo {volume} {134}},\ \bibinfo {pages} {183402} (\bibinfo {year} {2025})}\BibitemShut {NoStop}%
\bibitem [{\citenamefont {de~Jongh}\ \emph {et~al.}(2025)\citenamefont {de~Jongh}, \citenamefont {Verstraten}, \citenamefont {Dixmerias}, \citenamefont {Daix}, \citenamefont {Peaudecerf},\ and\ \citenamefont {Yefsah}}]{deJongh2025}%
  \BibitemOpen
  \bibfield  {author} {\bibinfo {author} {\bibfnamefont {T.}~\bibnamefont {de~Jongh}}, \bibinfo {author} {\bibfnamefont {J.}~\bibnamefont {Verstraten}}, \bibinfo {author} {\bibfnamefont {M.}~\bibnamefont {Dixmerias}}, \bibinfo {author} {\bibfnamefont {C.}~\bibnamefont {Daix}}, \bibinfo {author} {\bibfnamefont {B.}~\bibnamefont {Peaudecerf}},\ and\ \bibinfo {author} {\bibfnamefont {T.}~\bibnamefont {Yefsah}},\ }\bibfield  {title} {\bibinfo {title} {Quantum gas microscopy of fermions in the continuum},\ }\href {https://doi.org/10.1103/PhysRevLett.134.183403} {\bibfield  {journal} {\bibinfo  {journal} {Physical Review Letters}\ }\textbf {\bibinfo {volume} {134}},\ \bibinfo {pages} {183403} (\bibinfo {year} {2025})}\BibitemShut {NoStop}%
\bibitem [{\citenamefont {Xiang}\ \emph {et~al.}(2025)\citenamefont {Xiang}, \citenamefont {Cruz-Colón}, \citenamefont {Chua}, \citenamefont {Milner}, \citenamefont {de~Hond}, \citenamefont {Fricke},\ and\ \citenamefont {Ketterle}}]{Xiang2025}%
  \BibitemOpen
  \bibfield  {author} {\bibinfo {author} {\bibfnamefont {J.}~\bibnamefont {Xiang}}, \bibinfo {author} {\bibfnamefont {E.}~\bibnamefont {Cruz-Colón}}, \bibinfo {author} {\bibfnamefont {C.~C.}\ \bibnamefont {Chua}}, \bibinfo {author} {\bibfnamefont {W.~R.}\ \bibnamefont {Milner}}, \bibinfo {author} {\bibfnamefont {J.}~\bibnamefont {de~Hond}}, \bibinfo {author} {\bibfnamefont {J.~F.}\ \bibnamefont {Fricke}},\ and\ \bibinfo {author} {\bibfnamefont {W.}~\bibnamefont {Ketterle}},\ }\bibfield  {title} {\bibinfo {title} {In situ imaging of the thermal de {B}roglie wavelength in an ultracold {B}ose gas},\ }\href {https://doi.org/10.1103/PhysRevLett.134.183401} {\bibfield  {journal} {\bibinfo  {journal} {Physical Review Letters}\ }\textbf {\bibinfo {volume} {134}},\ \bibinfo {pages} {183401} (\bibinfo {year} {2025})}\BibitemShut {NoStop}%
\bibitem [{\citenamefont {Torquato}(2018)}]{Torquato2018}%
  \BibitemOpen
  \bibfield  {author} {\bibinfo {author} {\bibfnamefont {S.}~\bibnamefont {Torquato}},\ }\bibfield  {title} {\bibinfo {title} {Hyperuniform states of matter},\ }\href {https://doi.org/https://doi.org/10.1016/j.physrep.2018.03.001} {\bibfield  {journal} {\bibinfo  {journal} {Physics Reports}\ }\textbf {\bibinfo {volume} {745}},\ \bibinfo {pages} {1} (\bibinfo {year} {2018})},\ \bibinfo {note} {hyperuniform States of Matter}\BibitemShut {NoStop}%
\bibitem [{\citenamefont {Anderson}(1967)}]{Anderson1967}%
  \BibitemOpen
  \bibfield  {author} {\bibinfo {author} {\bibfnamefont {P.~W.}\ \bibnamefont {Anderson}},\ }\bibfield  {title} {\bibinfo {title} {Infrared catastrophe in {F}ermi gases with local scattering potentials},\ }\href {https://doi.org/10.1103/PhysRevLett.18.1049} {\bibfield  {journal} {\bibinfo  {journal} {Phys. Rev. Lett.}\ }\textbf {\bibinfo {volume} {18}},\ \bibinfo {pages} {1049} (\bibinfo {year} {1967})}\BibitemShut {NoStop}%
\bibitem [{\citenamefont {Fey}\ \emph {et~al.}(2020)\citenamefont {Fey}, \citenamefont {Schmelcher}, \citenamefont {Imamoglu},\ and\ \citenamefont {Schmidt}}]{Fey2020}%
  \BibitemOpen
  \bibfield  {author} {\bibinfo {author} {\bibfnamefont {C.}~\bibnamefont {Fey}}, \bibinfo {author} {\bibfnamefont {P.}~\bibnamefont {Schmelcher}}, \bibinfo {author} {\bibfnamefont {A.}~\bibnamefont {Imamoglu}},\ and\ \bibinfo {author} {\bibfnamefont {R.}~\bibnamefont {Schmidt}},\ }\bibfield  {title} {\bibinfo {title} {Theory of exciton-electron scattering in atomically thin semiconductors},\ }\href {https://doi.org/10.1103/PhysRevB.101.195417} {\bibfield  {journal} {\bibinfo  {journal} {Phys. Rev. B}\ }\textbf {\bibinfo {volume} {101}},\ \bibinfo {pages} {195417} (\bibinfo {year} {2020})}\BibitemShut {NoStop}%
\bibitem [{\citenamefont {Z{\"u}rn}(2013)}]{Zurn2013}%
  \BibitemOpen
  \bibfield  {author} {\bibinfo {author} {\bibfnamefont {G.~e.~a.}\ \bibnamefont {Z{\"u}rn}},\ }\bibfield  {title} {\bibinfo {title} {Precise characterization of $^6${L}i {F}eshbach resonances using trap-sideband-resolved rf spectroscopy of weakly bound molecules},\ }\href@noop {} {\bibfield  {journal} {\bibinfo  {journal} {Phys. Rev. Lett.}\ }\textbf {\bibinfo {volume} {110}},\ \bibinfo {pages} {135301} (\bibinfo {year} {2013})}\BibitemShut {NoStop}%
\bibitem [{\citenamefont {Ketterle}\ and\ \citenamefont {Zwierlein}(2008)}]{ketterle2008}%
  \BibitemOpen
  \bibfield  {author} {\bibinfo {author} {\bibfnamefont {W.}~\bibnamefont {Ketterle}}\ and\ \bibinfo {author} {\bibfnamefont {M.~W.}\ \bibnamefont {Zwierlein}},\ }\bibfield  {title} {\bibinfo {title} {Making, probing and understanding ultracold {F}ermi gases},\ }\href@noop {} {\bibfield  {journal} {\bibinfo  {journal} {La Rivista del Nuovo Cimento}\ }\textbf {\bibinfo {volume} {31}},\ \bibinfo {pages} {247} (\bibinfo {year} {2008})}\BibitemShut {NoStop}%
\bibitem [{\citenamefont {Wade}\ \emph {et~al.}(2011)\citenamefont {Wade}, \citenamefont {Baillie},\ and\ \citenamefont {Blakie}}]{Wade2011}%
  \BibitemOpen
  \bibfield  {author} {\bibinfo {author} {\bibfnamefont {A.~C.~J.}\ \bibnamefont {Wade}}, \bibinfo {author} {\bibfnamefont {D.}~\bibnamefont {Baillie}},\ and\ \bibinfo {author} {\bibfnamefont {P.~B.}\ \bibnamefont {Blakie}},\ }\bibfield  {title} {\bibinfo {title} {Direct simulation {M}onte-{C}arlo method for cold-atom dynamics: Classical {B}oltzmann equation in the quantum collision regime},\ }\href {https://doi.org/10.1103/PhysRevA.84.023612} {\bibfield  {journal} {\bibinfo  {journal} {Phys. Rev. A}\ }\textbf {\bibinfo {volume} {84}},\ \bibinfo {pages} {023612} (\bibinfo {year} {2011})}\BibitemShut {NoStop}%
\bibitem [{\citenamefont {Sykes}\ and\ \citenamefont {Bohn}(2015)}]{Sykes2015}%
  \BibitemOpen
  \bibfield  {author} {\bibinfo {author} {\bibfnamefont {A.~G.}\ \bibnamefont {Sykes}}\ and\ \bibinfo {author} {\bibfnamefont {J.~L.}\ \bibnamefont {Bohn}},\ }\bibfield  {title} {\bibinfo {title} {Nonequilibrium dynamics of an ultracold dipolar gas},\ }\href {https://doi.org/10.1103/PhysRevA.91.013625} {\bibfield  {journal} {\bibinfo  {journal} {Phys. Rev. A}\ }\textbf {\bibinfo {volume} {91}},\ \bibinfo {pages} {013625} (\bibinfo {year} {2015})}\BibitemShut {NoStop}%
\bibitem [{\citenamefont {Wigner}(1955)}]{Wigner1955}%
  \BibitemOpen
  \bibfield  {author} {\bibinfo {author} {\bibfnamefont {E.~P.}\ \bibnamefont {Wigner}},\ }\bibfield  {title} {\bibinfo {title} {Lower limit for the energy derivative of the scattering phase shift},\ }\href {https://doi.org/10.1103/PhysRev.98.145} {\bibfield  {journal} {\bibinfo  {journal} {Phys. Rev.}\ }\textbf {\bibinfo {volume} {98}},\ \bibinfo {pages} {145} (\bibinfo {year} {1955})}\BibitemShut {NoStop}%
\bibitem [{\citenamefont {Journeaux}\ \emph {et~al.}(2025)\citenamefont {Journeaux}, \citenamefont {Veschambre}, \citenamefont {Lecomte}, \citenamefont {Uzan}, \citenamefont {Dalibard}, \citenamefont {Werner}, \citenamefont {Petrov},\ and\ \citenamefont {Lopes}}]{journeaux2025}%
  \BibitemOpen
  \bibfield  {author} {\bibinfo {author} {\bibfnamefont {A.}~\bibnamefont {Journeaux}}, \bibinfo {author} {\bibfnamefont {J.}~\bibnamefont {Veschambre}}, \bibinfo {author} {\bibfnamefont {M.}~\bibnamefont {Lecomte}}, \bibinfo {author} {\bibfnamefont {E.}~\bibnamefont {Uzan}}, \bibinfo {author} {\bibfnamefont {J.}~\bibnamefont {Dalibard}}, \bibinfo {author} {\bibfnamefont {F.}~\bibnamefont {Werner}}, \bibinfo {author} {\bibfnamefont {D.~S.}\ \bibnamefont {Petrov}},\ and\ \bibinfo {author} {\bibfnamefont {R.}~\bibnamefont {Lopes}},\ }\href {https://arxiv.org/abs/2509.03567} {\bibinfo {title} {Two-body contact dynamics in a {B}ose gas near a {F}ano-{F}eshbach resonance}} (\bibinfo {year} {2025}),\ \Eprint {https://arxiv.org/abs/2509.03567} {arXiv:2509.03567 [cond-mat.quant-gas]} \BibitemShut {NoStop}%
\bibitem [{\citenamefont {Skipetrov}(2016)}]{Skipetrov2016PRB}%
  \BibitemOpen
  \bibfield  {author} {\bibinfo {author} {\bibfnamefont {S.~E.}\ \bibnamefont {Skipetrov}},\ }\bibfield  {title} {\bibinfo {title} {Finite-size scaling analysis of localization transition for scalar waves in a three-dimensional ensemble of resonant point scatterers},\ }\href {https://doi.org/10.1103/PhysRevB.94.064202} {\bibfield  {journal} {\bibinfo  {journal} {Phys. Rev. B}\ }\textbf {\bibinfo {volume} {94}},\ \bibinfo {pages} {064202} (\bibinfo {year} {2016})}\BibitemShut {NoStop}%
\end{thebibliography}

%

\end{document}